\documentclass[a4paper,11pt]{article}
\pdfoutput=1 

\usepackage{jheppub}
\usepackage{color} 
\usepackage{float}
\usepackage{ulem}

\usepackage{amsmath,graphicx,amssymb,subfigure,bbm}
\usepackage[all]{xy}

\def\beq{\begin{equation}}
\def\eeq{\end{equation}}
\def\beqa{\begin{eqnarray}}
\def\eeqa{\end{eqnarray}}

\def\R{ {\mathbb R} }

\begin{document}

\title{Magnetising the ${\cal N}=4$ Super Yang-Mills plasma}

\author[a]{Alfonso Ballon-Bayona,}
\author[b,c,d]{Jonathan P. Shock}
\author[e,f]{and Dimitrios Zoakos}
\affiliation[a]{Instituto de F\'{i}sica, Universidade
Federal do Rio de Janeiro, \\
Caixa Postal 68528, RJ 21941-972, Brazil.}       
\affiliation[b]{Department of Mathematics and Applied Mathematics,
University of Cape Town, \\
Private Bag, Rondebosch 7700, South Africa.}
\affiliation[c]{The National Institute for Theoretical and Computational Sciences,
Private Bag X1, Matieland, South Africa.}
\affiliation[d]{Institut National de la Recherche Scientifique, Centre \'Energie, Mat\'eriaux et T\'el\'ecommunications, 1650 Boul. Lionel Boulet, Varennes, Qu\'ebec J3X 1S2, Canada}
\affiliation[e]{Department of Physics, 
National and Kapodistrian University of Athens, 15784 Athens, Greece.}
\affiliation[f]{Department of Engineering and Informatics, 
Hellenic American University, \\ 
436 Amherst st, Nashua, NH 03063 USA}

\emailAdd{aballonb@if.ufrj.br}
\emailAdd{jonathan.shock@uct.ac.za}
\emailAdd{zoakos@gmail.com}

\abstract{We investigate the thermodynamics of the anisotropic magnetic $\rm AdS_5$ black brane solution found by D'Hoker and Kraus \cite{DHoker:2009mmn}. This solution is the gravity dual of  a strongly coupled ${\cal N}=4$ Super Yang-Mills plasma in $\R^{3,1}$, with temperature $T$, in the presence of a magnetic field ${\cal B}$. Following the procedure of holographic renormalisation we calculate the Gibbs free energy and the holographic stress tensor of the conformal plasma.  
We evaluate several thermodynamic quantities including the magnetisation, the anisotropic pressures and the speeds of sound. Our results are consistent with an RG flow from a perturbed $\rm AdS_5$ black brane at small ${\cal B}/T^2$ to a $\rm \bf BTZ \times \R^2$ black brane at large ${\cal B}/T^2$. We also perform a phenomenological analysis where we compare the thermodynamics of a magnetised conformal plasma against the lattice QCD results for the thermodynamics of the magnetised quark-gluon plasma.}

\maketitle

\flushbottom

\section{Introduction}

The description of macroscopic properties of strongly coupled matter is a challenging problem due to the necessity of non-perturbative methods. In the case of the quark-gluon plasma (QGP), produced in heavy ion collisions at the  Relativistic Heavy Ion Collider (RHIC) and the 
Large Hadron Collider (LHC), it is important to reach a proper understanding of the thermodynamic and hydrodynamic properties of Quantum Chromodynamics (QCD) at temperatures where non-perturbative effects are relevant.

A very interesting theoretical framework for understanding strongly coupled matter is the AdS/CFT (Anti-de-Sitter/Conformal Field Theory) correspondence and more generally the gauge/gravity duality (for a recent book see for example \cite{Ammon:2015wua}). According to the AdS/CFT correspondence, a strongly coupled fluid with conformal symmetry at finite temperature in $d$ dimensions is mapped to a $d+1$ dimensional asymptotically AdS black brane. In the case of $d=4$ very precise predictions were done for the strongly coupled conformal plasma formed in the ${\cal N}=4$ Super Yang-Mills theory in the large $N_c$ limit \cite{Gubser:1996de,Burgess:1999vb,Policastro:2001yc,Policastro:2002se}. The latter is the maximal supersymmetric extension of 4d Yang-Mills theory. Although this theory is quite different from real QCD, some of the macroscopic properties at finite temperature were found to be close to those of real QCD. The most striking example was the prediction of the shear viscosity to entropy density ratio \cite{Policastro:2001yc,Policastro:2002se} which is very close to the expected result for the quark-gluon plasma observed in heavy ion collisions. 

In this work we will describe some thermodynamic and hydrodynamics properties of the strongly coupled ${\cal N}=4$ Super Yang-Mills plasma in the presence of a magnetic field. We revisit the magnetic black brane solution found by D'Hoker and Kraus \cite{DHoker:2009mmn,DHoker:2009ixq} in order to provide a systematic description of the thermodynamic properties of this magnetised conformal plasma, focusing on the anisotropic effects caused by the magnetic field. From a theoretical point of view, a non-zero magnetic field is considered a useful method for investigating non-perturbative aspects of QCD \cite{Kharzeev:2013jha}. From an experimental point of view, anisotropic effects in the quark-gluon plasma are expected when intense magnetic fields, of the order of  $e B /\Lambda^2_{QCD} \sim 5 -10$,  are produced by the spectator nuclei in off-central heavy ion collisions \cite{Skokov:2009qp}. Even in the absence of magnetic fields it is expected that anisotropic effects should play an important role in the description of the quark-gluon plasma soon after the collision \cite{Fukushima:2016xgg}.
Lastly, an interesting prediction for the magnetised quark-gluon plasma is the so-called chiral magnetic effect, associated with topological charge changing transitions \cite{Fukushima:2008xe}. This effect has not yet been observed in heavy ion collisions  because there are anisotropies in the directions  transverse to the magnetic field, that produce similar effects that are difficult to disentangle. Since the description of anisotropic effects in the QCD quark-gluon plasma is a very hard problem due to the non-perturbative behaviour of QCD at strong coupling, investigating this problem in the strongly coupled ${\cal N}=4$ Super Yang-Mills plasma can provide very useful insights. 

Anisotropic effects in $3+1$ dimensional strongly coupled fluids have been previously investigated in AdS/CFT \cite{DHoker:2009mmn,DHoker:2009ixq,DHoker:2010onp,Mateos:2011ix,Mateos:2011tv,Rebhan:2011vd,Chernicoff:2012iq,Giataganas:2012zy,Chernicoff:2012gu,Critelli:2014kra,Rougemont:2014efa,Ammon:2020rvg} and holographic models for QCD  \cite{Ammon:2012qs,Conde:2016hbg,Penin:2017lqt, Jokela:2019tsb,Patino:2012py,Jain:2015txa,Finazzo:2016mhm,Gursoy:2016ofp,Giataganas:2017koz,Gursoy:2018ydr,Arefeva:2018hyo,Braga:2018zlu,Avila:2018hsi,Avila:2019pua,Bohra:2019ebj,Avila:2020ved,Rougemont:2020had,Gursoy:2020kjd,Arefeva:2020vae}.
An interesting consequence of anisotropy is the violation of the shear viscosity bound in the strongly coupled plasma \cite{Rebhan:2011vd,Critelli:2014kra,Jain:2015txa}. There are two other interesting phenomena associated with the presence of a magnetic field that have been investigated in holographic models for QCD. Magnetic catalysis, which refers to the enhancement of chiral symmetry breaking due to the magnetic field, originally discovered in effective field theory models \cite{Gusynin:1994re,Gusynin:1994xp} and confirmed in QCD  (for a review see \cite{Miransky:2015ava}), was investigated extensively in holography \cite{Filev:2007gb,Bergman:2008sg,Johnson:2008vna,Filev:2009xp,Filev:2011mt,Erdmenger:2011bw,Jokela:2013qya,He:2020fdi,Ballon-Bayona:2020xtf}. Inverse magnetic catalysis, which refers to the enhancement of chiral restoration due to the magnetic field, was initially discovered at finite chemical potential \cite{Inagaki:2003yi} and also described in holographic QCD \cite{Preis:2010cq,Preis:2012fh,Ballon-Bayona:2017dvv}. The staggering observation of inverse magnetic catalysis at zero chemical potential in lattice QCD \cite{Bali:2011qj, Bali:2012zg,Bruckmann:2013oba}  motivated further effort in holographic QCD \cite{Ballon-Bayona:2013cta,Mamo:2015dea,Rougemont:2015oea,Dudal:2015wfn,Dudal:2016joz,Evans:2016jzo,Gursoy:2016ofp,Li:2016gfn,Gursoy:2017wzz,Rodrigues:2017iqi,Rodrigues:2018pep,Gursoy:2020kjd}. For a recent book and review on holographic QCD models in the presence of a magnetic field see \cite{Gursoy:2021kqt} and \cite{Gursoy:2021efc} respectively. Finally, the study of magnetic field effects in $2+1$ dimensional strongly coupled fluids via the AdS/CFT correspondence and the gauge gravity/duality has a very rich history starting with the pioneering works of \cite{Hartnoll:2007ai,Hartnoll:2007ip}.

The outline of this paper is as follows. In section \ref{Sec:CFT} we describe some thermodynamic and hydrodynamic properties of a conformal fluid in the presence of a magnetic field. In particular, we derive the equation of state of a magnetised conformal plasma and useful relations for the components of the stress-energy tensor. In section \ref{Sec:GravityDual} we describe the gravity dual of a 4d magnetised ${\cal N}=4$ Super Yang-Mills plasma in terms of the asympotically $AdS_5$ magnetic black brane solution found by D'Hoker and Kraus.  We implement a holographic renormalisation procedure that allows us to obtain a Gibbs free energy consistent with the thermodynamics of a conformal plasma and a thermodynamic entropy consistent with the Bekenstein-Hawking entropy. 

Moreover, our results for the Gibbs free energy will be consistent with an RG flow from a perturbed $\rm AdS_5$ black brane at small ${\cal B}/T^2$ to a $\rm \bf BTZ \times \R^2$ black brane at large ${\cal B}/T^2$. This is shown in section \ref{Sec:Numerics} where we numerically evaluate several thermodynamic quantities including the magnetisation, anisotropic pressures and the speeds of sound. We finish that section with a phenomenological analysis where the thermodynamic results of a strongly coupled magnetised conformal plasma are compared against the lattice QCD results for the quark-gluon plasma obtained in \cite{Bali:2014kia}. We also describe the two analytical solutions found in the regimes of small ${\cal B}/T^2$and large ${\cal B}/T^2$. The perturbed $\rm AdS_5$ black brane solution at small ${\cal B}/T^2$ is described in appendix \ref{Appendix:smallb} whilst the $\rm \bf BTZ \times \R^2$ black brane at large ${\cal B}/T^2$ is described in appendix \ref{Appendix:BTZ}. Details of the calculation of the holographic stress tensor are given in appendix \ref{Appendix:stress-tensor}.


\section{The magnetised conformal plasma}
\label{Sec:CFT}

In this section we describe both thermodynamic and hydrodynamic properties of a conformal fluid in the presence of a magnetic field. We will follow \cite{Caldarelli:2008ze} and describe the equation of state and the stress-energy tensor of the fluid. From there we will derive some universal relations for the components of the stress tensor that will prove very useful when investigating the holographic stress tensor dual to the 5d metric describing the magnetic black brane. 

\subsection{Equation of state}

We start with the Gibbs free energy 
\begin{equation}
G = E - T S = G(T,V,{\cal B}) \, , \label{GibbsEn}
\end{equation}
where $E$ is the magnetic enthalpy \cite{Carlin:1986,CASTELLANO2003146}, related to the internal energy $U$ and magnetisation $M$ by  $E = U - M {\cal B}$. 
We are working in the grand canonical ensemble where the thermodynamic variables are ${\cal B}$, $T$ and $V$ and the relevant thermodynamic potential is the Gibbs free energy \footnote{In the canonical ensemble the thermodynamic variables would be $M$, $T$ and $V$ and the relevant thermodynamic potential would be the Helmoltz free energy $F=U-TS$.}. 
From conformal symmetry and extensivity, the Gibbs free energy of a conformal fluid in $D$ dimensions takes the form
\begin{equation}
G =  V T^D \, g( b) \, , \label{GibbsFreeEn}
\end{equation}
where $b= {\cal B}/T^2$ is the dimensionless ratio of the magnetic field and the temperature squared.
Under scaling symmetry $x \to \alpha x$, we have 
\begin{equation}
T \to \alpha^{-1} T \, , \, V \to \alpha^{D-1} V \, , {\cal B} \to \alpha^{-2} {\cal B} \, , \, G \to \alpha^{-1} G  \,. 
\end{equation} 
Defining $\lambda \equiv \alpha^{-1}$, the transformation rule for the Gibbs free energy can be written as 
\begin{equation}
G (\lambda T , \lambda^{1-D} V , \lambda^2 {\cal B}) = \lambda \, G (T,V,{\cal B}) \,. \label{GibbsTransf}
\end{equation}
We denote the rescaled quantities as
\begin{equation}
T' \equiv \lambda T  \, , \, V' \equiv \lambda^{1-D} V \, , 
{\cal B}' \equiv \lambda^2 {\cal B} \, , G' \equiv G(T',V',{\cal B}') \,.  
\end{equation}
Differentiating \eqref{GibbsTransf} with respect to $\lambda$ and setting $\lambda=1$ we obtain
\begin{align}
G = \Big [\frac{d G'}{d \lambda} \Big ]_{\lambda=1} &= \Big [ 
\frac{\partial G'}{\partial T'} \frac{\partial T'}{\partial \lambda} +  \frac{\partial G'}{\partial V'} \frac{\partial V'}{\partial \lambda} + \frac{\partial G'}{\partial {\cal B}'} \frac{\partial {\cal B}'}{\partial \lambda}\Big ]_{\lambda=1} \nonumber \\
&= \Big [ - S' T - P' (1-D) \lambda^{-D} V - V M (2 \lambda) {\cal B} \Big ]_{\lambda=1} \nonumber \\
&= - T S - (1-D) P V - 2 V M {\cal B}  \, , \label{CFTRel}
\end{align}
where we used the definitions of entropy $S$, pressure $P$ and magnetisation density $M$ given by
\begin{equation}
S = - \frac{\partial G}{\partial T} \quad , \quad 
P = - \frac{\partial G}{\partial V} = - \frac{G}{V} \quad , \quad 
 M = - \frac{1}{V} \frac{\partial G}{\partial {\cal B}} \,,
\end{equation}
Combining \eqref{GibbsEn} and \eqref{CFTRel} we find the equation of state for the magnetic enthalpy:
\begin{equation}
E = (D-1) P V - 2 V M {\cal B} \, .
\end{equation}
Since $P V = -G$, the CFT identity \eqref{CFTRel} can be written as
\begin{equation}
D \, G = - T S - 2 V M {\cal B} \,. 
\end{equation}
For the specific case of $D=4$ we obtain the thermodynamic relations
\begin{align}
G &= - \frac14 T S - \frac12 V M {\cal B} = - P V \, , \nonumber \\
E &= G + T S = \frac34 T S - \frac12 V M {\cal B} 
= 3 P V - 2 V M {\cal B} \,. \label{EOSDeq4}
\end{align}
For fixed volume $V$, it is convenient to define the densities
\begin{equation}
{\cal G} = \frac{G}{V} \quad , \quad {\cal F} = \frac{F}{V} \quad , \quad 
{\rho} = \frac{E}{V} \quad , \quad 
{\cal U} = \frac{U}{V} \quad , \quad 
{\cal S} = \frac{S}{V}  \,. \label{thermodqts}
\end{equation}
We are able to write these quantities by very simple relations in terms of the dimensionless Gibbs free energy density
\begin{equation}
g(b)=\frac{{\cal G}}{T^4}   \,,
\end{equation}
defined previously in \eqref{GibbsFreeEn}, with $b={\cal B}/T^2$. 
The thermodynamic quantities in \eqref{thermodqts} become
\begin{align}
    \text{Entropy Density:} \,\,\,\,\,\,{\cal S} &= - \frac{\partial {\cal G} }{\partial T} =  T^3 \Big [ 2 b \, g'(b) - 4  g(b)\Big ] = -2T^3b^3\partial_b\left(g/b^2\right)\,\nonumber\\
\text{Magnetisation Density:} \,\,\,\,\,\,M &= - \frac{\partial {\cal G}}{\partial {\cal B}}   
= - T^2 g'(b)
\nonumber\\
\text{Helmholtz free energy Density:} \,\,\,\,\,\,{\cal F} &= {\cal G} + M B 
 \nonumber \\
&= T^4 \Big [ g(b) - b g'(b) \Big ] = -T^4b^2\partial_b\left(g/b\right)
\nonumber\\
\text{Internal energy density:}\,\,\,\,\,\,{\cal U} &= {\cal F} + T {\cal S} \nonumber \\
&= T^4 \Big [ b g'(b) - 3 g(b) \Big ] =   T^4 b^4 \partial_b\left(g/b^3\right)
\nonumber \\
\text{Magnetic enthalpy density:}\,\,\,\,\,\,\rho &= {\cal U} - M {\cal B} = {\cal G} + T {\cal S} \nonumber \\
&= T^4 \Big [ 2 b g'(b) - 3 g(b) \Big ] = 2T^4 b^\frac{5}{2}\partial_b\left(g/b^\frac{3}{2}\right)
\end{align}
where the Gibbs free energy, entropy and magnetisation densities satisfy the conformal identity \begin{eqnarray}
{\cal G} = - \frac14 T {\cal S} - \frac12 M {\cal B} \,. 
\end{eqnarray}

From the magnetisation we can extract the magnetic susceptibility
\begin{align}
\chi &= \frac{\partial M}{\partial {\cal B}} = - T^2 g''(b) \frac{\partial b}{\partial {\cal B}} = - g''(b) \, , 
\end{align}
and the pyro-magnetic coefficient
\begin{align}
\xi &= \frac{\partial M}{\partial T} = - 2 T g'(b) - T^2 g''(b) \frac{\partial b}{\partial T} \nonumber \\
&=2  T  \Big [  b g''(b) - g'(b)  \Big ] \,. 
\end{align}
These quantities are related by the conformal identity 
\begin{equation} \label{conformal_identity}
M = \chi {\cal B} + \frac12 \xi T \,.  
\end{equation}
Finally, we obtain the specific heat 
\begin{equation}
C_{V,{\cal B}} = \frac{ \partial \rho}{ \partial T}
= T^3 \Big [ -12 g(b) + 10 b g'(b) - 4 b^2 g''(b) \Big ] \,.
\end{equation}
Note that 
\begin{equation}
C_{V,{\cal B}} = T \frac{ \partial {\cal S} }{ \partial T}   = {\cal S}
\frac{ \partial \ln {\cal S} }{ \partial \ln T} \,.
\end{equation}
We can thus calculate all thermodynamic quantities in terms of derivatives of the dimensionless Gibbs free energy density $g(b)$. This will prove very useful later for our numerical calculations.

\subsection{Stress tensor}
In order to calculate the speed of sound and the hydrodynamic pressures, it is necessary to calculate the stress-energy tensor of the fluid. This can be written as 
\begin{equation} \label{magstress}
T^{\mu \nu} = (\rho + P) u^{\mu}u^{\nu} + P \eta^{\mu \nu} 
- {\cal M}^{\mu \rho} {\cal F}^{\nu}_{\, \rho} \, ,
\end{equation}
where $\rho$ is the magnetic enthalpy density and $P$ is the thermodynamic pressure, both defined at thermodynamic equilibrium. We remind the reader that we are working in the grand canonical ensemble where the thermodynamic variables are ${\cal B}$, $T$ and $V$, the relevant thermodynamic potential is the Gibbs free energy $G$ and the thermodynamic pressure is given by $P= -G/V$. 

The last term in \eqref{magstress} represents the coupling between the electromagnetic field strength and the polarisation tensor. The latter is defined by 
\begin{equation}
{\cal M}^{\mu \rho} \equiv - \frac{1}{V} \frac{\partial G}{\partial {\cal F}_{\mu \rho}} \,. 
\end{equation}
For a 4d conformal fluid in the presence of a magnetic field in the $z$ direction, the only non-zero components of the polarisation tensor are
\begin{equation}
{\cal M}^{12} = - {\cal M}^{21} = - \frac{1}{V} \frac{\partial G}{\partial {\cal B}} =  M \, .
\end{equation}
In the rest-frame of the fluid  $u^{\mu}=(1,0,0,0)$, so the anisotropic stress tensor becomes
\begin{equation}
T^{\mu \nu} = {\rm diag} (\rho, P_x, P_x , P_z ) \, ,
\end{equation}
where
\begin{align}
\rho &=   {\cal G} + T {\cal S}  = 
 \frac34 T {\cal S} - \frac12 M {\cal B} \, , \nonumber \\
P_x &= P - M {\cal B}  = \frac14 T {\cal S} - \frac12 M {\cal B} \, , \nonumber \\
P_z &= P = - {\cal G}  = \frac14 T {\cal S} + \frac12 M {\cal B}  \label{Pressures}
\end{align}
where ${\cal G}$ and ${\cal S}$ represent the Gibbs free energy density and entropy density respectively. The trace of the stress tensor is given by $-\rho + 2 P_x + P_z$ and vanishes as a consequence of conformal symmetry.  

Note that we distinguish between the hydrodynamic pressures $P_x$ and $P_z$ and the thermodynamic pressure $P$. It turns out that the hydrodynamic pressure parallel to the magnetic field is equal to the thermodynamic pressure, i.e. $P_z=P$. Since we are working in the grand canonical ensemble we further identify $P_z$ with $-{\cal G}$ with ${\cal G}$ the Gibbs free energy density. Note also that the hydrodynamic pressure transverse to the magnetic field is identified with minus the Helmoltz free energy density, i.e.  $P_x = P - M {\cal B} = - {\cal F}$. 

If we consider sound propagation in the $x$ and $z$ directions we have two different values for the squared speed of sound:
\begin{align}
c_{s,x}^2 =  \left ( \frac{ \partial P_x}{ \partial {\cal \rho}}\right )_{\cal B} = \frac{ {\cal S} - \xi {\cal B}}{C_{V,{\cal B}}}  \quad , \quad
c_{s,z}^2 = \left ( \frac{ \partial P_z}{ \partial {\cal \rho}}\right )_{\cal B} = \frac{{\cal S}}{C_{V,{\cal B}}} \,. 
\end{align} 
Note that 
\begin{equation}
 \frac{1}{c_{s,z}^2} = \frac{C_{V,{\cal B}}}{ \cal S} = \frac{ \partial \ln {\cal S}}{ \partial \ln T} \,.     
\end{equation}
This concludes the derivation of all hydro and thermodynamic quantities which we will be able to calculate for the magnetic plasma.


\section{Gravity dual of the magnetised conformal plasma}
\label{Sec:GravityDual}

In this section we will present the gravity solution that interpolates between an $AdS_5$ space on the boundary 
and a $\rm \bf BTZ \times \R^2$ black hole in the deep IR. This is a solution that was first investigated by D'Hoker and Kraus in 
\cite{DHoker:2009mmn} (with a very interesting extension after the addition of an electric charge density in 
\cite{DHoker:2009ixq, DHoker:2010onp}). 
However, in contrast with \cite{DHoker:2009ixq}, we will introduce a diffeomorphism invariant counterterm that leads naturally to a Gibbs free energy and a stress tensor consistent with a conformal fluid.  We will also discuss the non-diffeomorphism invariant counterterm of \cite{DHoker:2009ixq} and show how it generates a conformal anomaly.

After the holographic renormalisation procedure we will further subtract the action at zero temperature in order to ensure that the end of the RG flow is the BTZ solution. Moreover, the subtracted thermodynamic and hydrodynamic quantities will be equivalent for the diffeomorphism invariant renormalisation proposed in this work and the non-diffeomorphism invariant considered in  \cite{DHoker:2009ixq}.  In particular, the subtracted stress tensor will be traceless, as expected for a conformal plasma.
This will be analysed in full detail in the rest of this section.


\subsection{The asymptotically $AdS_5$ magnetic black brane}

The Einstein-Maxwell action in five dimensions with a negative cosmological constant is given by the following expression 
\begin{equation} \label{EMaction5d}
S = \sigma \int d^5 x \sqrt{-g} \Bigg [ R + \frac{12}{\ell^2} - \frac{c}{4}  \, F_{mn}F^{mn} \Bigg ] 
\quad {\rm with} \quad \sigma=\frac{1}{16 \, \pi \, G_5}
\end{equation}
where the five dimensional Newton constant $G_5$ is fixed by the AdS/CFT dictionary to be $G_5=(\pi/2) \ell^3 N_c^{-2}$. 
The gauge field coupling is set to $c=4 \, \ell^2$ and in the following we work in units where $\ell=1$. 
The Einstein-Maxwell equations are
\begin{equation} \label{EinstMaxEq}
R_{mn} - \frac{R}{2} \, g_{mn} - 6 \, g_{mn} \, = \, 2  \, T_{mn}  
\quad \& \quad 
\nabla_m F^{mn} \, = \, 0
\end{equation}
where
\begin{equation}
T_{mn} = F_{mp} F^{\, \, p}_{n} - \frac{1}{4} \, g_{mn} \, F_{pq} \, F^{pq}
\end{equation}
is the five dimensional stress energy tensor associated with the gauge field. 
Note that taking the trace of the first equation in \eqref{EinstMaxEq} leads to the following form for the Ricci scalar
\begin{equation}  \label{Ricciscalar}
R \, = \, - \, 20 - \frac43 \, T \ = \, - \, 20  + \, \frac13  \, F_{pq} F^{pq} \, .
\end{equation}

The aim is to find an asymptotically $AdS_5$ black brane solution dual to an ${\cal N}=4$ Super Yang-Mills plasma in the 
presence of a magnetic field. Denoting the boundary coordinates by $(t,x,y,z)$ and choosing $z$ as the direction of the 
magnetic field, we look for a metric that preserves the $SO(2)$ symmetry in the $(x,y)$ plane. A suitable ansatz for the metric 
and field strength can be written as \cite{DHoker:2009mmn}
\begin{equation} \label{ansatz}
ds^2  =   - U(r) dt^2  + \frac{dr^2}{U(r)} + e^{2V(r)} \left ( d x^2 + d y^2 \right ) + e^{2 W(r)} dz^2 
\quad \& \quad  
F = B \, dx \wedge d y \, .
\end{equation}
The five dimensional radial coordinate $r$ goes from the horizon radius $r=r_h$ (where $U(r_h)=0$) to the 
boundary at $r \to \infty$.
As noted in \cite{DHoker:2009mmn}, the magnetic field $B$ in the Einstein Maxwell ansatz has to be rescaled as ${\cal B}= \sqrt{3}B$ 
in order to define the correct parameter on the ${\cal N}=4$ SYM side.
For the numerical analysis, it is convenient to define the dimensionless radial coordinate $\tilde r \equiv r/r_h$ 
and rescale the fields and coordinates as follows
\begin{equation} \label{rescaled-r}
U(r) \equiv r_h^2 \, \tilde U (\tilde r) \,  , \quad  
\left( e^{V(r)} , e^{W(r)} \right)  \equiv r_h \left( e^{\tilde V (\tilde r)} , e^{\tilde W (\tilde r)} \right )\, , \quad 
B \equiv r_h^2 \, \tilde B 
\quad \& \quad  
x_{\mu} \equiv \frac{\tilde {x}_{\mu}}{r_h} \,. 
\end{equation}
Note that the horizon radius is  located at $\tilde r=1$. The ansatz \eqref{ansatz} keeps the same form in terms of the rescaled fields and coordinates. Plugging this ansatz into \eqref{EinstMaxEq} we find that the nonzero components of the Einstein-Maxwell equations take the form
\begin{eqnarray} \label{fieldeqsv2} 
{\tilde U}\, \left( \tilde{V}''-{\tilde W}''\right) + \Big[ {\tilde U}' +  {\tilde U}\, \left(2 \,  \tilde{V}'+{\tilde W}'\right) \Big] \, 
\left( \tilde{V}'-{\tilde W}'\right) & = & - 2\,  {\tilde B}^2\, e^{-4\,  \tilde{V}}
\nonumber \\
2 \,  \tilde{V}'' + {\tilde W}'' + 2\, \left({\tilde V}'\right)^2  + \left({\tilde W}'\right)^2 & = & 0
\nonumber \\
\frac{1}{2} \, {\tilde U}'' + \frac{1}{2} {\tilde U}' \, \left(2 \,  \tilde{V}' + {\tilde W}'\right) & =  &
 4 + \frac{2}{3} \, {\tilde B}^2\, e^{-4\,  \tilde{V}}
\nonumber \\
2\, {\tilde U}' \,  \tilde{V}'+{\tilde U}' \, {\tilde W}' + 2 \, {\tilde U} \, ( \tilde{V}')^2 + 4 \, {\tilde U} \,  \tilde{V}' \, {\tilde W}'
& = &  12  - 2 \, {\tilde B}^2\, e^{-4\,  \tilde{V}}
\end{eqnarray}
where $'$ represents $d/d \tilde r$. The first three differential equations are dynamical whereas the last differential equation 
is a constraint. One of the dynamical equations can be omitted when solving the system, 
since it can be obtained from the others. We exclude the final dynamical equation and include the constraint when solving numerically.


\subsubsection{Near horizon and near boundary asymptotics}

Near the horizon $\tilde r=1$, regularity implies that the solutions to the field equations 
\eqref{fieldeqsv2} admit a Taylor expansions and take the form
\begin{align}  \label{IRNLO}
\tilde U( \tilde r) &= \tilde{U}_{h,1} \, \left(\tilde{r}-1\right)+\left[\frac{5 \, {\tilde B}^2}{3 \, \tilde{v}_{h,0}^4}-2\right]
\,\left(\tilde{r}-1\right)^2 + {\cal O} (\tilde r-1)^3
 \nonumber \\
\frac{e^{\tilde V(\tilde r)}}{\tilde v_{h,0}} &= 1 - \frac43 \frac{\tilde B^2 - 3 \, \tilde v_{h,0}^4}{ \tilde U_{h,1} \, 
\tilde v_{h,0}^4}\, \left(\tilde r -1\right)  
-\frac{4}{9} \, \frac{\tilde{B}^2-3 \, \tilde{v}_{h,0}^4}{\tilde{U}_{h,1}^2 \, \tilde{v}_{h,0}^8} \,\tilde B^2\,  \left(\tilde{r}-1\right)^2
+ {\cal O} (\tilde r-1)^3
\\
\frac{e^{\tilde W (\tilde r)}}{\tilde w_{h,0}} &= 1 + \frac23 \frac{\tilde B^2 + 6 \, \tilde v_{h,0}^4}{ \tilde U_{h,1} \, \tilde v_{h,0}^4}\,  
\left(\tilde r-1\right) + 
\frac{8}{9} \, \frac{\tilde{B}^2-3 \, \tilde{v}_{h,0}^4}{\tilde{U}_{h,1}^2 \, \tilde{v}_{h,0}^8}\,\tilde B^2 \,  
\left(\tilde{r}-1\right)^2+ {\cal O} (\tilde r-1)^3 \, .
\nonumber
\end{align}
The near horizon solutions are characterised by the three parameters $\tilde U_{h,1}$, $\tilde v_{h,0}$ and $\tilde w_{h,0}$.  The parameters  $\tilde v_{h,0}$ and $\tilde w_{h,0}$ are the coefficients in the near horizon expansion for the fields $\tilde v = \exp (\tilde V) $ and $\tilde w = \exp (\tilde W) $. For simplicity, we  choose coordinates such that $\tilde v_{h,0}=1$, i.e. $\tilde V(\tilde r=1)=0$. The remaining two parameters $\tilde U_{h,1}$ and $\tilde w_{h,0}$ will be the initial data for the numerical solutions to the field equations \eqref{fieldeqsv2}. These parameters will be determined as a function of $\tilde B$ by imposing the AdS boundary conditions 
$\tilde V(\tilde r \to \infty) = \tilde W(\tilde r \to \infty) = \ln \tilde r$.\footnote{The boundary condition $\tilde U (\tilde r \to \infty) = \tilde r^2$ is automatically satisfied.} To summarise, we initially have five integration constants in the Einstein-Maxwell equations. Regularity at the horizon reduces these to three: $\tilde U_{h,1}, \tilde v_{h,0}$ and $\tilde w_{h,0}$. We also have the freedom to redefine the $x$ and $y$ coordinates allowing us to set the integration constant $\tilde v_{h,0}$ to $1$ and the remaining two constants are obtained numerically by imposing the AdS asymptotic form  at large $\tilde r$.

Near the boundary the asymptotic solutions to the field equations \eqref{fieldeqsv2} take the form 
\begin{align} \label{UVNLO}
\frac{\tilde U(\tilde r)}{\tilde r^2} &= 1  + \tilde U_{\infty,1} \,  \tilde r^{-1} + \frac{\tilde U_{\infty,1}^2}{4} \, \tilde r^{-2}
- \frac23 \, \tilde B^2 \, \tilde r^{-4} \, \ln \tilde r + \tilde U_{\infty,4} \, \tilde r^{-4} 
\nonumber \\
& + \frac23 \, \tilde B^2 \, \tilde U_{\infty,1} \, \tilde r^{-5} \, \ln \tilde r
- \frac13 \, \tilde U_{\infty,1} \, \left(\tilde B^2 + 3 \, \tilde U_{\infty,4} \right) \, \tilde r^{-5}
+ {\cal O}(\tilde r^{-6} \, \ln \tilde r) + {\cal O}(\tilde r^{-6})  
 \nonumber \\
\frac{e^{\tilde V(\tilde r)}}{\tilde r} &=  1 + \frac{ \tilde U_{\infty,1}}{2} \, \tilde r^{-1} + \frac16 \, \tilde B^2 \, \tilde r^{-4} \, 
\ln \tilde r +  \tilde v_{\infty,4} \, r^{-4}   
\nonumber \\
&- \frac14 \, \tilde B^2 \, \tilde U_{\infty,1} \, \tilde r^{-5} \, \ln \tilde r 
+ \frac{1}{12} \, \tilde U_{\infty,1} \, \left(\tilde B^2 - 18 \, \tilde v_{\infty,4}\right) \, \tilde r^{-5}
+ {\cal O}(\tilde r^{-6} \, \ln \tilde r) +  {\cal O}(\tilde r^{-6})
 \nonumber \\
\frac{e^{\tilde W (\tilde r)}}{\tilde r} &= 1 + \frac{\tilde U_{\infty,1}}{2} \, \tilde r^{-1} - \frac13 \, \tilde B^2 \, \tilde r^{-4} \, 
\ln \tilde r -  2 \, \tilde v_{\infty,4} \, r^{-4}   
\nonumber \\
&+  \frac12 \, \tilde B^2 \, \tilde U_{\infty,1} \, \tilde r^{-5} \, \ln \tilde r
- \frac16 \, \tilde U_{\infty,1} \, \left(\tilde B^2 - 18 \, \tilde v_{\infty,4}\right) \, \tilde r^{-5}
+ {\cal O}(\tilde r^{-6} \, \ln \tilde r) +  {\cal O}(\tilde r^{-6})\, .
\end{align}
The three UV parameters $\tilde U_{\infty,1}$, $\tilde U_{\infty,4}$ and $\tilde v_{\infty,4}$ will be extracted from the numerical solutions and are all functions of $\tilde{B}$. We will see later that the UV parameters  $\tilde U_{\infty,4}$ and $\tilde v_{\infty,4}$ will be associated with the 
time and space components of the dual stress tensor describing the four dimensional conformal fluid in the presence of a magnetic field.


\subsubsection{Temperature and entropy density}

\begin{figure}[ht]
\centering
\includegraphics[width=6cm]{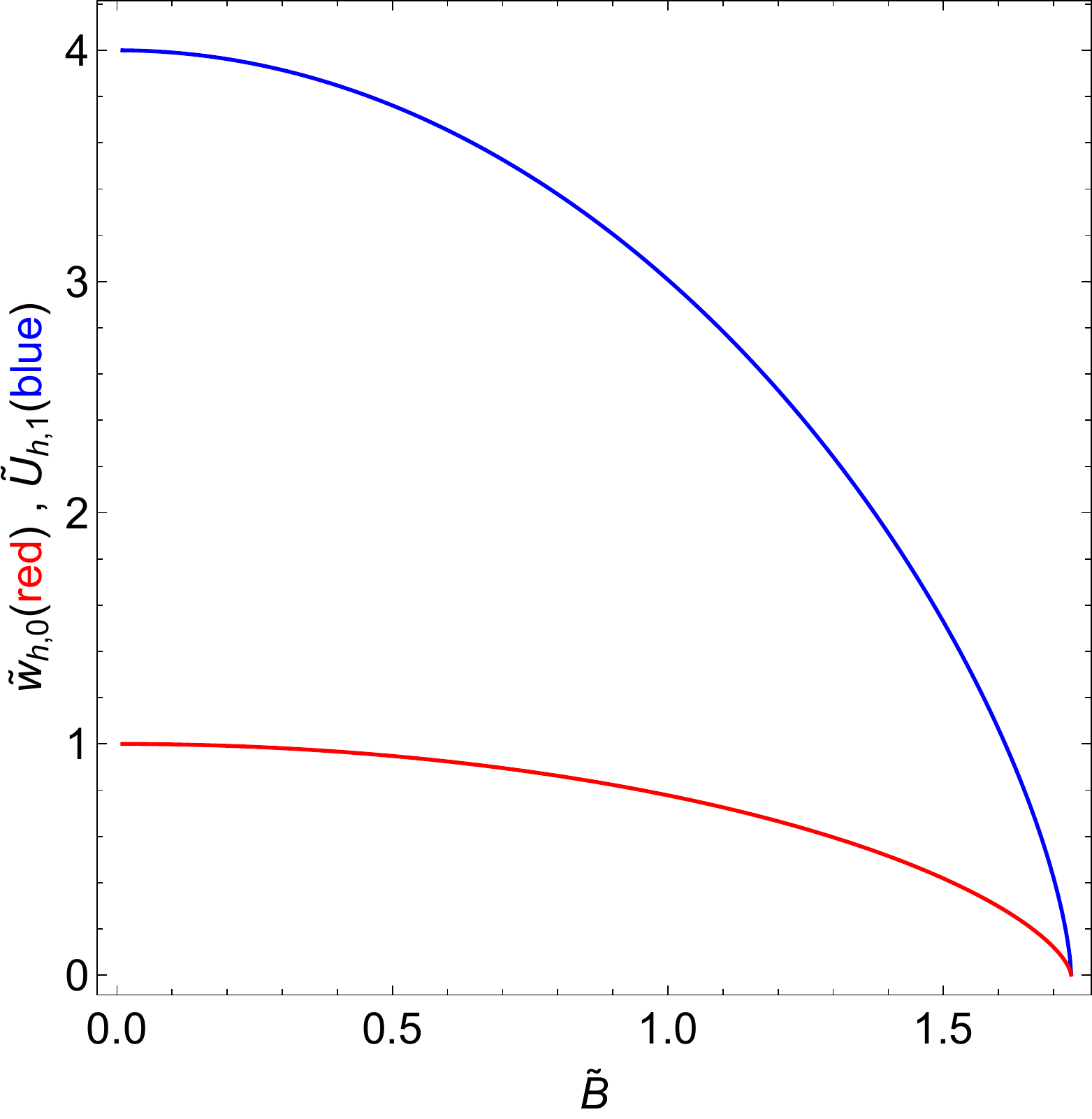}
\hspace{1cm}
\includegraphics[width=6.5cm]{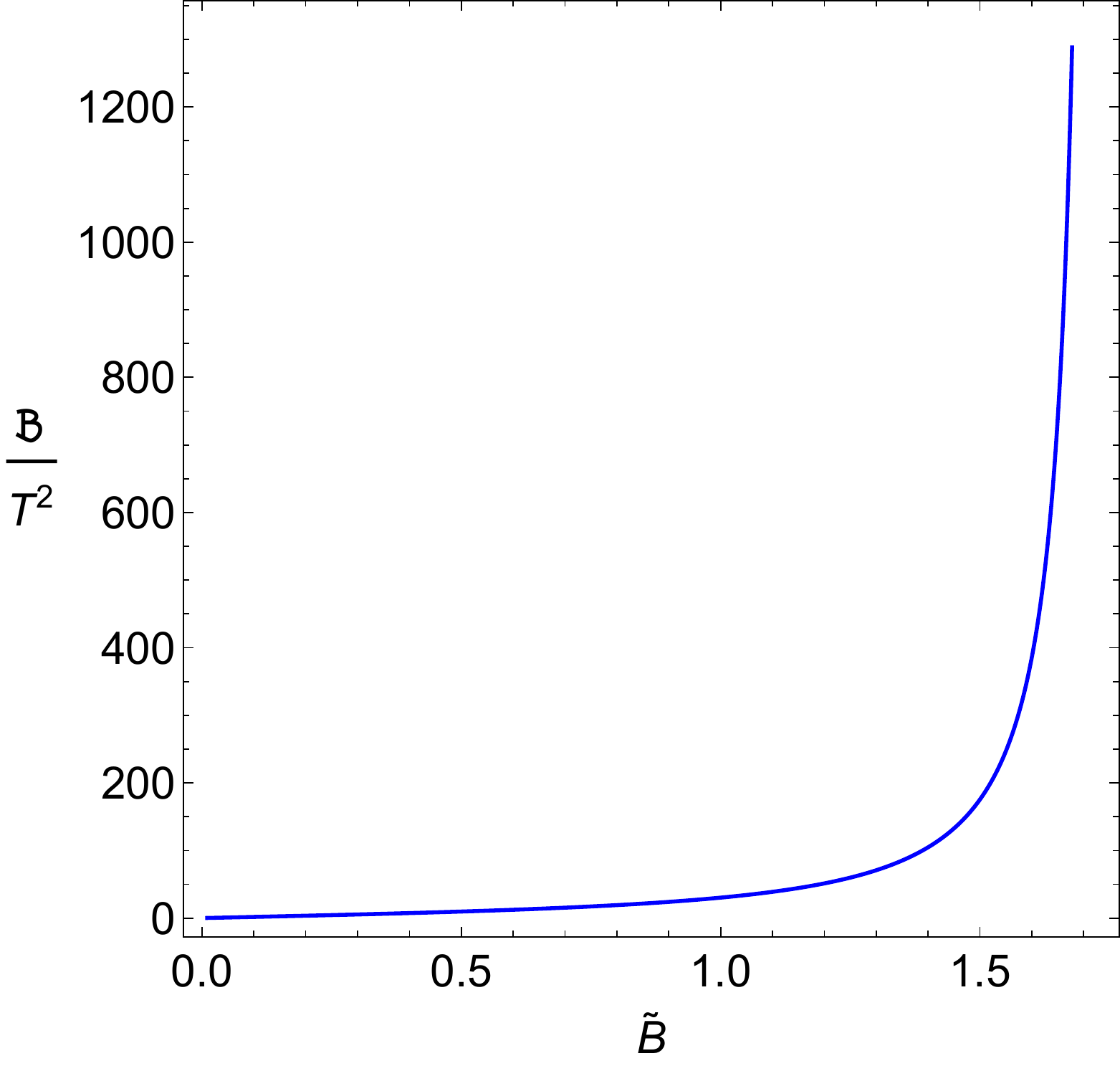}
\caption{Left panel: The horizon parameters $\tilde U_{h,1}$ and $\tilde w_{h,0}$ as functions of $\tilde B$. 
Notice that both  $\tilde U_{h,1}$ and $\tilde w_{h,0}$ are defined in units where the horizon radius is at $\tilde r=1$.
Right panel: the dimensionless ratio ${\cal B}/T^2$ as a function of $\tilde B$.}
\label{Fig:Uh1wh0vsB}
\end{figure}

The temperature is fixed in terms of the horizon parameters by requiring the absence of a conical singularity at $r=r_h$ . This is given by the well-known formula
\begin{equation} \label{temperature}
T \, = \, \frac{U'(r_h)}{4 \, \pi} 
\quad \Rightarrow \quad 
T \, =  \, \frac{r_h}{4 \, \pi} \, \tilde U_{h,1} 
\end{equation}
The entropy density is related to the horizon area $A_h$ by the Bekenstein-Hawking formula 
\begin{equation} \label{entropy}
{\cal S} \equiv \frac{S}{V_3} = \frac{A_h}{ 4 \, G_5 \, V_3}  = 4 \, \pi \, \sigma \,  e^{2 \, V(r_h)+ W(r_h)} 
\quad \Rightarrow \quad 
{\cal S} = 4 \, \pi \, \sigma \, r_h^3 \, \tilde v_{h,0}^2 \, \tilde w_{h,0} 
\end{equation}
where $V_3 = \int \int \int  dx  dy dz$ is the three dimensional spatial volume. 
It is convenient to define the dimensionless ratio of the magnetic field and the temperature as follows
\begin{equation} \label{BovT2}
\frac{{\cal B}}{T^2} \, = \, 16 \, \sqrt{3} \, \pi^2 \, \frac{\tilde B}{\tilde U_{h,1}^2} \, .  
\end{equation}
As a reminder, we have chosen coordinates in such a way that $\tilde v_{h,0}=1$ and the numerical solution 
is characterised by the initial data $\tilde U_{h,1}$ and  $\tilde w_{h,0}$, at the horizon $\tilde r=1$. 
These parameters are determined as functions of $\tilde B$ from the boundary conditions $\tilde V(\tilde r \to \infty) = \tilde W(\tilde r \to \infty) = \ln \tilde r$ and in  figure \ref{Fig:Uh1wh0vsB} we present the result of the numerical evaluation. 
Notice that for small values of $\tilde B$, $\tilde w_{h,0}$ is close to 1 and the background is isotropic. As $\tilde B$ increases
$\tilde w_{h,0}$ deviates from 1 (which is the value of $\tilde v_{h,0}$ for any $\tilde B$ ) and the background 
becomes anisotropic. The more we increase  $\tilde B$, the more anisotropic the background becomes. 

Using \eqref{BovT2}, on the right panel of figure \ref{Fig:Uh1wh0vsB}, we present the dimensionless ratio 
${\cal B}/T^2$ as a function of $\tilde B$ and we find a monotonically increasing behaviour. 
In particular, the limit $\tilde B \to \sqrt{3}$ corresponds to the limit ${\cal B} \to \infty$ (very strong magnetic fields) or $T \to 0$ (very low temperatures). This last observation can also be seen from the plot of $\tilde U_{h,1}$ as a function of $\tilde B$ in the 
left panel of  figure \ref{Fig:Uh1wh0vsB}. More precisely, when $\tilde B \to \sqrt{3}$ then  $\tilde U_{h,1} \to 0$ and the 
temperature goes to zero.

The dimensionless ratio ${\cal S}/T^3$ of the entropy density and the temperature can be obtained using 
\eqref{temperature} and \eqref{entropy}, but we will display our results in the next section, where we recover the 
entropy formula also from the holographic calculation of the Gibbs free energy.


\subsection{Thermodynamics from holography}

An alternative and in principle complementary way to calculate the entropy of a gravity background is by using holographic renormalisation in order to calculate the free energy from the Euclidean on-shell action. This is the path we are going to follow in this subsection 
and we will use the (Gibbs) free energy to study the thermodynamics. In the main text we present the case for 
a general magnetic field, whilst in appendix \ref{Appendix:smallb} we elaborate on the case of small $B/T^2$ using a perturbative 
solution. 


\subsubsection{The renormalised on-shell action}

The Euclidean version of the renormalised Einstein-Maxwell action in \eqref{EMaction5d} can be written as follows
\begin{equation} \label{ren-Action}
S_{ren} = S_{{\cal M}} + S_{\partial {\cal M}} + S_{ct}
\end{equation}
where
\begin{equation} \label{EMBulk5d}
S_{{\cal M}} \, = \,  - \, \sigma \int_{ {\cal M}} d^5 x \sqrt{g}  \left ( R + 12  - F_{mn}^2 \right )  =   
\sigma \int_{ {\cal M}} d^5 x \sqrt{g} \left( 8  +  \frac{2}{3}  \,F_{mn}^2 \right)
\end{equation}
is the Einstein-Maxwell bulk term, 
\begin{equation}  \label{GK5d}
S_{ \partial {\cal M}} \, =  \, - \, 2 \, \sigma \int_{ \partial {\cal M}} 
d^4 x \sqrt{\gamma} \, K
\end{equation}
is the Gibbons-Hawking boundary term and the counter-term action $S_{ct}$ will be determined later. 
Note that to obtain \eqref{EMBulk5d} we have plugged the expression \eqref{Ricciscalar} for the Ricci scalar. 
In \eqref{GK5d} the quantity $K$ is the trace of the extrinsic tensor $K_{\mu \nu}$ and $\gamma$ is the determinant of 
the boundary induced metric $\gamma_{\mu \nu}$.  

Plugging the ansatz \eqref{ansatz} into the bulk term \eqref{EMBulk5d} and using \eqref{rescaled-r} we obtain
\begin{equation} \label{EMBulk5dv2}
S_{ {\cal M}} \, =  \sigma \, V_3 \, \beta \, r_h^4 \int d \tilde r \,  e^{2 \tilde V + \tilde W} 
\left[  8  + \frac43 \tilde B^2 e^{-4 \, \tilde V} \right]
\end{equation}
where $\beta$ is the period of the imaginary time $\tau = i \, t$, related to the temperature by $\beta=T^{-1}$.
To further simplify \eqref{EMBulk5dv2}, we use the third Einstein equation in \eqref{fieldeqsv2} 
to write the integrand as a total derivative
\begin{equation} \label{EMBulk5dv3}
S_{ {\cal M}} \, = \,  \sigma \, V_3 \, \beta \, r_h^4 \int_{1}^{ \tilde r_0} d \tilde r \,  \partial_{\tilde r} 
\left[e^{2 \, \tilde V + \tilde W} \tilde U' \right] \, 
= \, \sigma \, V_3 \, \beta \, r_h^4  \, \Big[e^{2 \, \tilde V + \tilde W} \tilde U' \Big]_{\tilde r = 1}^{\tilde r = \tilde r_0}
\end{equation}
where $\tilde r = \tilde r_0$ and $\tilde r=1 $ are the radial positions of the boundary and horizon respectively. 
In the end we will take the limit $\tilde r_0 \to \infty$. 
The trace of the extrinsic tensor can be written as 
\begin{equation}
K \, = \, \nabla_m \eta^m = \frac{1}{\sqrt{g}} \partial_m \left (\sqrt{g} \, \eta^m \right) 
\quad {\rm with} \quad 
\eta^m = \left( \sqrt{U(r)} , 0,0,0,0\right)  \,.  \label{eta}
\end{equation}
Using these results the Gibbons-Hawking term  \eqref{GK5d} takes the form
\begin{equation}  \label{GK5dv2}
S_{\partial {\cal M}}\,  =\, \sigma \, V_3 \, \beta \, r_h^4 
\left [ e^{2 \, \tilde V + \tilde W} \, \tilde U' - 2 \left ( e^{2 \, \tilde V + \tilde W} \tilde U \right )' \,  \right ]_{\tilde r = \tilde r_0} \,. 
\end{equation}
Summing \eqref{EMBulk5dv3} and \eqref{GK5dv2} we obtain the following expression for the first two terms in \eqref{ren-Action}
\begin{equation}  \label{OnShell5d}
S_{{\cal M}} + S_{\partial {\cal M}} 
 \, = \, - \,  \sigma \, V_3 \, \beta \, r_h^4 \, \Bigg \{  2 \Big [ \tilde U \Big ( e^{2 \, \tilde V+ \tilde W} \Big )' \,\Big ]_{\tilde r =\tilde r_0} 
 + \Big [ e^{2 \, \tilde V + \tilde W} \tilde U' \Big ]_{\tilde r = 1} \Bigg \} \,. 
\end{equation}
Plugging the asymptotic solutions \eqref{IRNLO} and \eqref{UVNLO} into the on-shell action \eqref{OnShell5d} we obtain 
\begin{align}  \label{OnShell5dv2}
S_{{\cal M}} + S_{\partial {\cal M}} 
&= \, - \, \sigma \, V_3 \, \beta \, r_h^4 \Bigg \{ 6 \Bigg [ \tilde r_0^4 + 2 \, \tilde U_{\infty,1} \, \tilde r_0^3 + 
\frac{3}{2} \, U_{\infty,1}^2 \, \tilde r_0^2 +  \frac{1}{2} \, \tilde U_{\infty,1}^3 \, \tilde r_0 
\nonumber \\
& - \frac{2}{3} \, \tilde B^2 \, \ln \tilde r_0 + \frac{1}{16} \, \tilde U_{\infty,1}^4 +  \tilde U_{\infty,4} \Bigg]+ 
\tilde U_{h,1} \tilde v_{h,0}^2 \tilde w_{h,0}  \Bigg \} \,.
\end{align}
Note that the UV parameter $\tilde U_{\infty,1}$ leads to multiple UV divergences for the on-shell action.  However, we can perform the renormalisation procedure for non-zero values of $\tilde U_{\infty,1}$ as we will show below. 
In this work we consider a diffemorphism-invariant  counterterm action given by
\begin{align} \label{counterterm-action}
S_{ct} &= \, \sigma \, \int d^4 x \sqrt{\gamma}  \, \Big [ a_1 + a_2  \, F_{\mu  \nu} F^{\mu \nu} \ln 
\left(F_{\mu  \nu} F^{\mu \nu}\right) + a_3  \, F_{\mu  \nu} F^{\mu \nu} \Big ]  
\nonumber \\
&= \sigma \, V_3 \, \beta \, r_h^4 \Bigg \{ a_1 \Bigg [ \tilde r_0^4 + 2 \, \tilde U_{\infty,1}\,  \tilde r_0^3 + 
\frac32 \, \tilde U_{\infty,1}^2 \, \tilde r_0^2 + \frac12 \, \tilde U_{\infty,1}^3 \, \tilde r_0   - 
\frac13 \, \tilde B^2 \, \ln \tilde r_0 
\nonumber \\
& + \, \frac{1}{16} \, \tilde U_{\infty,1}^4 + \frac12 \, \tilde U_{\infty,4} 
 \Bigg ] + 2  \, \tilde B^2 \, a_2 \Big [  \ln \left(2 \tilde B^2\right)  - 4 \ln \tilde r_0    \Big ] + 2   \tilde B^2 a_3   \Bigg \}
\end{align}
where we have used the asymptotic expansions \eqref{IRNLO} and \eqref{UVNLO}. 
In order to cancel the power-law and logarithmic  divergences in \eqref{OnShell5dv2} we choose $a_1=6$ and 
$a_2=\frac{1}{4}$. The value of $a_3$ is not fixed. The renormalised action  \eqref{ren-Action} reduces to
\begin{equation}  \label{ren-Action_v2}
S_{ren} \, = \, - \, \sigma \, V_3 \, \beta \, r_h^4 \Bigg [ 3 \, \tilde U_{\infty,4} 
+ \tilde U_{h,1} \, \tilde v_{h,0}^2 \, \tilde w_{h,0} - \tilde B^2 \, \ln \tilde B   
- \left (2 a_3 + \frac12 \, \ln 2\right) \tilde B^2 \Bigg ] \,.
\end{equation}
Notice that the renormalised action in \eqref{ren-Action_v2} depends both on UV data (through $\tilde U_{\infty,4} $) 
and IR data (through $\tilde U_{h,1}$, $\tilde v_{h,0}$ and $\tilde w_{h,0} $) but it is also scheme dependent, 
since we have not fixed the parameter $a_3$. 

Our holographic renormalisation procedure differs from previous approaches, for example \cite{DHoker:2009ixq}, by the logarithmic term that brings a non-trivial dependence on the magnetic field.  The diffeomorphism-invariant counterterm in \eqref{counterterm-action} does not break conformal invariance and leads naturally to a Gibbs free energy and a stress tensor consistent with the thermodynamics and hydrodynamics of a conformal plasma described in section \ref{Sec:CFT}. We will describe later in this section that the non-diffeomorphism invariant counterterm considered in \cite{DHoker:2009ixq} can be thought as a deformation of the counterterm in \eqref{counterterm-action} that leads to the breaking of conformal invariance and the emergence of a conformal anomaly for the stress tensor.

\subsubsection{The Gibbs free energy density}
\label{subsubsec:Gibbs}

The renormalised Gibbs free energy density takes the form
\begin{equation} \label{FreeEn}
{\cal G} = \frac{T \, {\cal S}_{ren} }{V_3} \, = \, - \, \sigma \, r_h^4 \, \Bigg [ 3 \, \tilde U_{\infty,4} 
+ \tilde U_{h,1}\,  \tilde v_{h,0}^2 \, \tilde w_{h,0}  - \tilde B^2 \, \ln \tilde B - \left(2 \, a_3 + \frac{ \ln 2}{2}\right) \tilde B^2 \Bigg ] 
\equiv   r_h^4  \, \tilde {\cal G} \,.
\end{equation}
We remind the reader that $\sigma=1/(16 \pi G_5)=N_c^2/(8 \pi^2)$
and that $r_h$ and $\tilde B$ are related by $ r_h^2 \tilde B = B = {\cal B}/\sqrt{3}$.  Note that the last two terms in \eqref{FreeEn} depend only on ${\cal B}^2$. This ${\cal B}^2$ contribution to the free energy is scheme dependent and does not contribute to the entropy density.   

The magnetic enthalpy density can be written as $\rho = r_h^4 \tilde \rho$ with 
\begin{equation}
\tilde \rho \, = \, \tilde {\cal G} + \tilde T \, \tilde {\cal S} \, =  \, \frac{N_c^2}{8 \,\pi^2} \, \Bigg[- \, 3 \, \tilde U_{\infty,4} + \tilde B^2 \, \ln \tilde B + 
\left(2 \, a_3 + \frac{1}{2} \ln 2\right) \tilde B^2 \Bigg]\, ,
\end{equation}
and $\tilde T$ and $\tilde {\cal S}$ are defined by the relations $T = r_h \tilde T$, ${\cal S} = r_h^3 \tilde {\cal S}$. 

At this point we can calculate various thermodynamic quantities analytically in the cases of zero magnetic field and zero temperature.
 
 
\paragraph{Zero magnetic field:}

In the case of $B=0$, we find the following analytic solution
\begin{equation}
e^{\tilde V} =  \tilde r \, , \quad 
e^{\tilde W} =  \tilde r \quad \&  \quad 
\tilde U = \tilde r^2 \left ( 1 - \tilde r^{-4} \right )
\end{equation}
from which we can easily extract the quantities that enter in \eqref{FreeEn}, namely
\begin{equation}
\tilde U_{\infty,4} = - 1 \, , \quad 
\tilde v_{h,0} = \tilde w_{h,0} = 1 \quad \& \quad 
\tilde U_{h,1} = 4 \, .
\end{equation}
Temperature, entropy density and free energy density take the form
\begin{equation}
T_{B=0} = \frac{r_h}{\pi} \, ,  \quad  
{\cal S}_{B=0} = \frac{\pi^2}{2} N_c^2 T^3
\quad \& \quad
{\cal G}_{B=0} =  - \frac{\pi^2}{8} N_c^2  T^4 \, . 
\end{equation}
Our results for the entropy and free energy densities of  the strongly coupled ${\cal N}=4$ super Yang-Mills plasma in the limit of zero magnetic field reduces to those obtained in the pioneer work \cite{Gubser:1996de}. 
These quantities satisfy the following thermodynamic relations
\begin{equation}
{\cal G}_{B=0} = - \frac14 (T {\cal S})_{B=0} \quad \& \quad 
\rho_{B=0} = \frac34 (T {\cal S})_{B=0} \, .  
\end{equation}


\paragraph{Zero temperature:}

Solving the equations of motion numerically, we know that when the dimensionless parameter $\tilde B$ 
approaches $\sqrt{3}$, $B\rightarrow\infty$ and the value of $U_{h,1}$ approaches zero, and this is identified as the zero temperature limit. Indeed as there is only a single dimensionless parameter, $\frac{B}{T^2}$, the limit $B\rightarrow\infty$ is equivalent to $T\rightarrow 0$.
Putting all this information together we arrive to the following zero temperature (numerical) solution
\begin{equation} \label{zero_T_sol}
\tilde B_{T=0} = \sqrt{3} \, , \quad 
\tilde U_{h,1}^{T=0} = 0 \, , \quad
\tilde U_{\infty,4}^{T=0} = \tilde U_{\infty,4}(\sqrt{3}) \quad \& \quad 
r_h^{T=0} = \sqrt{ \frac{B}{\sqrt{3}}} = \sqrt{ \frac{{\cal B}}{3}}
\end{equation}
where we have used \eqref{rescaled-r} to determine $r_h^{T=0} $ as a function of $B$ (and ${\cal B}$).
Substituting \eqref{zero_T_sol} in  \eqref{FreeEn} we have
\begin{equation} \label{free-energy-norm-zeroT}
{\cal G}_{T=0} \, = \, \rho_{T=0} \, = \, - \, \frac{\sigma}{3} \, {\cal B}^2 \Big [  \tilde U_{\infty,4}(\sqrt{3}) -  \ln (\sqrt{3}) \Big ] + 
\frac{\sigma}{3} \left(2 a_3 + \frac{\ln 2}{2} \right) {\cal B}^2  \,. 
\end{equation}
We can use this result to define a new version of the renormalised free energy, that is 
\begin{align}  \label{FreeEn_v2}
{\cal G}_r &\equiv {\cal G} - {\cal G}_{T=0} 
\nonumber \\
&= \, - \, \sigma \, r_h^4 \, \Big [ 3 \, \tilde U_{\infty,4}  + \tilde U_{h,1} \, \tilde v_{h,0}^2 \, \tilde w_{h,0}  
- \tilde B^2 \, \ln \tilde B \Big ] + 
\frac{\sigma}{3} \, {\cal B}^2 \, \Big [ \tilde U_{\infty,4}(\sqrt{3}) -  \ln (\sqrt{3}) \Big ] \,. 
\end{align}
The free energy as it is defined in \eqref{FreeEn_v2} has two important advantages with respect to the 
expression in \eqref{FreeEn}. The first one is that it is scheme independent since it does not depend on $a_3$ and the second is 
that it describes the RG flow between an $AdS_5$ background on the boundary and a  $\rm \bf BTZ \times \R^2$ black hole in the deep IR. 
Without the subtraction we propose in \eqref{FreeEn_v2}, in the IR we obtain the entropy of the  $\rm \bf BTZ \times \R^2$ solution (since
the derivative of the term that contains $a_3$ with respect to the temperature is zero) but not its magnetisation. 
The subtraction procedure of \eqref{FreeEn_v2} will effectively fix the value of $a_3$ to the value that is needed in order for the 
free energy to flow from $AdS_5$ to  $\rm \bf BTZ \times \R^2$. 

In practice we can evaluate the renormalised free energy as follows
\begin{equation}
{\cal G}_r =  r_h^4 \, \tilde {\cal G}(\tilde B) - \frac{1}{9} \, \tilde {\cal G}(\sqrt{3}) \,  {\cal B}^2
\end{equation}
where $r_h$ can be obtained for a given $T$  and ${\cal B}$ by the relation 
\begin{equation}
 T = r_h \tilde T (\tilde B)
\end{equation}
where $\tilde B$ is a function of $b={\cal B}/T^2$.  We will be interested in the subtracted version of the 
dimensionless free energy
\begin{equation} \label{grfromtG}
g_r(b) \equiv  \frac{{\cal G}_r}{T^4} = g(b) - \frac{1}{9} \, \tilde {\cal G}(\sqrt{3}) \, b^2
\quad {\rm where} \quad g(b) = \frac{ \tilde {\cal G}(\tilde B)}{ \left[\tilde T(\tilde B)\right]^4 }  \, .
\end{equation} 
This result explicitly shows that the subtracted renormalised Gibbs free energy density is consistent with conformal invariance because the dimensionless ratio ${\cal G}_r/T^4$ depends only on the dimensionless ratio $b= {\cal B}/T^2$.   The relation between ${\cal B}/T^2$ and the parameter $\tilde B$ was displayed on the right panel of figure \ref{Fig:Uh1wh0vsB}.


\subsection{The holographic stress tensor}
\label{subsec:Hologstress}

The 4d stress-energy tensor is given by the on-shell variation of the Einstein-Maxwell action, in Lorentzian signature, with respect to the source $\gamma_{\mu \nu}^{(0)}= r_0^{-2} \gamma_{\mu \nu}$
\begin{equation}
T^{\mu \nu}_{ren}  =  \frac{2}{\sqrt{-\gamma^{(0)}}} \frac{ \delta S_{ren}}{ \delta \gamma^{(0)}_{\mu \nu}} 
=   \frac{2 \, r_0^6}{\sqrt{-\gamma}} \frac{ \delta S_{ren}}{ \delta \gamma_{\mu \nu}}
= T^{\mu \nu}_{reg} + T^{\mu \nu}_{ct}
\end{equation}
where 
\begin{equation} \label{Tmunu_Reg+ct}
\frac{T^{\mu \nu}_{reg}}{r_0^{6}} =   \frac{2 }{\sqrt{-\gamma}} \frac{ \delta (S_{{\cal M}}+ 
S_{\partial {\cal M} })}{ \delta \gamma_{\mu \nu}} 
=  - \, 2  \, \sigma  \left( K^{\mu \nu} - K \gamma^{\mu \nu} \right) 
\quad \& \quad
\frac{T^{\mu \nu}_{ct}}{r_0^{6}} =  - \frac{2}{\sqrt{-\gamma}} \frac{ \delta S_{ct}}{\delta \gamma_{\mu \nu}} \, , 
\end{equation}
where $-S_{ct}$ is the Lorentzian version of the counter-term action \eqref{counterterm-action}. Here $K_{\mu \nu} = \nabla_{\mu} \eta_{\nu}$ is the extrinsic curvature tensor with $\eta^m$ given in \eqref{eta},  $K = \gamma^{\mu \nu} K_{\mu \nu}$ is the trace of the extrinsic curvature tensor and $\gamma_{\mu \nu}$ the induced metric that takes the form
\begin{equation}
\gamma_{\mu \nu} = {\rm diag} \Big ( - U \, , e^{2V} , e^{2V}, e^{2W}  \Big )_{r_0} \, .
\end{equation}    
Varying the counter-term action \eqref{counterterm-action} we obtain
\begin{eqnarray} \label{Tmunu_ct_v2}
&&\frac{T^{\mu \nu}_{ct}}{2 \, \sigma \, r_0^{6}} =  - \frac{a_1}{2}\,  \gamma^{\mu \nu} 
+ 2 \, a_2 \Big[ {\cal T}^{\mu \nu} \, \ln (F_{\rho  \sigma} F^{\rho \sigma}) +F^{\mu \rho} F^{\nu}_{\, \, \, \rho} \Big]
+ 2 \, a_3  {\cal T}^{\mu \nu}  \nonumber \\
&&
\quad \quad {\rm with} \quad 
{\cal T}^{\mu \nu} =  F^{\mu \rho} F^{\nu}_{\, \, \, \rho} - \frac{1}{4} \, F_{\rho  \sigma} F^{\rho \sigma} \gamma^{\mu \nu} \, ,
\end{eqnarray}
and we choose $a_1=6$ and $a_2=\frac{1}{4}$ as before in order to cancel the UV divergences. Substituting the ansatz \eqref{ansatz} in \eqref{Tmunu_Reg+ct} and \eqref{Tmunu_ct_v2} and using the
UV asymptotic behaviour \eqref{UVNLO}, together with the zero temperature subtraction that we introduced in the 
previous subsection, we arrive at the following expressions for the components of the stress-energy tensor
\begin{align} \label{EMT_rho_v2}
\rho &=  - \langle T^t_{\,\, t} \rangle_r = - \frac{3\, r_h^4}{8\, \pi^2} \Bigg [ \tilde U_{\infty,4} -\frac{\tilde B^2}{6}\left[2 \, \tilde U_{\infty,4}(\sqrt{3})  + 
\ln \left(\frac{\tilde B^2}{3}\right)\right]  \Bigg ] 
 \\  \label{EMT_P_x_v2}
P_x & =\langle T^1_{\,\, 1} \rangle_r= \langle T^2_{\,\, 2} \rangle_r= -
\frac{r_h^4}{8\, \pi^2} \Bigg [ \tilde U_{\infty,4} - 8 \, \tilde v_{\infty,4} - \frac{\tilde B^2}{2}\left[2 \, \tilde U_{\infty,4}(\sqrt{3})  + 
\ln \left(\frac{\tilde B^2}{3}\right)\right]  \Bigg ] 
\\  \label{EMT_P_z_v2}
P_z &=\langle T^3_{\,\, 3} \rangle_r= - \frac{r_h^4}{8\, \pi^2} \Bigg [ \tilde U_{\infty,4} + 16 \, \tilde v_{\infty,4} 
+ \frac{\tilde B^2}{2}\left[2 \, \tilde U_{\infty,4}(\sqrt{3})  + 
\ln \left(\frac{\tilde B^2}{3}\right)\right]  \Bigg ] \, . 
\end{align}
More details about the intermediate steps that lead to the equations \eqref{EMT_rho_v2}, \eqref{EMT_P_x_v2} \&  \eqref{EMT_P_z_v2} can be found in appendix \ref{Appendix:stress-tensor}.
Notice that by setting $\tilde U_{\infty,4}=-1$ (this is the value of $\tilde U_{\infty,4}$ in the limit of small $B$) to  \eqref{EMT_rho_v2} we obtain the expression 
for the energy density in  \eqref{small_B_SE_tt}.
It can be easily checked that the trace of the stress tensor vanishes, i.e. 
\begin{equation}
\langle T^{\mu}_{\, \, \mu} \rangle_r  \, = \,  - \, \rho + 2 \, P_x + P_z \, =  \, 0
\end{equation}
as expected for a conformal fluid. In the next section we will obtain the hydrodynamic pressures from the components of the holographic stress tensor. Using  the holographic dictionary \eqref{EMT_rho_v2} we will find that the components of the stress-energy tensor are consistent with conformal magnetohydrodynamics, in particular it will satisfy the thermodynamic relations \eqref{Pressures} found at the end of section \ref{Sec:CFT}. But before that we briefly discuss below how the deformation of the diffeomorphism invariant counterterm into a non-diffeomorphism invariant counterterm leads to conformal anomaly.


\subsection{The non-diffeomorphism invariant counterterm and the conformal anomaly}

We end this section with a brief discussion of the non-diffeomorphism invariant counterterm proposed in \cite{DHoker:2009ixq}. We will show that this leads to a non-zero contribution to the trace of the stress tensor and can be interpreted in terms of a conformal anomaly in the 4d field theory  \cite{Fuini:2015hba,Janiszewski:2015ura,Ghosh:2021naw}. We will find, however, that despite the presence of the conformal anomaly the subtracted quantities obtained in the previous subsections remain conformally invariant. 

The counterterm action considered in \cite{DHoker:2009ixq} can be written as 
\begin{align}
S_{ct}^{(2)} &= \sigma \int d^4 x \sqrt{\gamma}  \Big [ a_1 + a_2  F_{\mu  \nu} F^{\mu \nu} \ln (r_0^{-4}) + a_3  F_{\mu  \nu} F^{\mu \nu} \Big ] \,. \label{nondiffSct} 
\end{align}
The non-diffeomorphism invariant counterterm  $S_{ct}^{(2)}$ in  \eqref{nondiffSct} can be thought as a deformation of the  diffeomorphism invariant counterterm $S_{ct}$ in \eqref{counterterm-action}. Namely, we have 
\begin{equation}
   S_{ct}^{(2)} =  S_{ct} + \Delta S_{ct} \, ,
\end{equation}
where 
\begin{equation}
\Delta S_{ct}
= - \sigma \, a_2 \int d^4 x \sqrt{\gamma}     F_{\mu  \nu} F^{\mu \nu}  \Big [ \ln (F_{\mu  \nu} F^{\mu \nu} ) + 4 \ln r_0  \Big ] \,,  \label{finitect}
\end{equation}
where $a_2=1/4$ \footnote{The coefficient  $a_2$ is fixed in the same way when we renormalise the action using the counterterm action $S_{ct}$ or the counterterm action $S_{ct}^{(2)}.$}. This difference can be interpreted as a finite counterterm that leads to conformal symmetry breaking. Indeed, the finite counterterm \eqref{finitect} leads to the following contribution to the Gibbs free energy density:
\begin{align}
\Delta {\cal G} = \frac{1}{V_3} T \Delta  S_{ct} = - \frac12 \sigma B^2 \ln (2 B^2)\,. \label{DeltaG}
\end{align}
This term breaks  conformal symmetry because the dimensionless ratio $\Delta {\cal G}/T^4$ does not depend only on $b={\cal B}/T^2$. However, since the term in \eqref{DeltaG} depends only on the magnetic field it does not contribute to the subtracted free energy density ${\cal G}_r = {\cal G} - {\cal G}_{T=0}$, which is still given by \eqref{FreeEn_v2}
and consistent with conformal invariance. The subtraction of the zero temperature free energy is also a natural way to construct quantities that are scheme independent. 

Similarly, the Lorentzian version of the finite counterterm \eqref{finitect} leads to the following contribution to the stress-energy tensor:
\begin{equation} \label{DeltaTmunu}
\Delta \langle T^{\mu \nu} \rangle =  
   - \frac12 \sigma B^2 \, {\rm diag} \Big (  \ln (2 B^2) \,  , 
 2 + \ln (2B^2)\,  , 2 + \ln (2B^2) \, ,  - \ln (2B^2)  \Big ) \,. 
\end{equation}
This in turn leads to a non-zero contribution to the trace
\begin{eqnarray}
\Delta \langle T^{\mu}_{\, \, \mu} \rangle = - 2 \sigma B^2 \,, 
\end{eqnarray}
which agrees with previous results for the 4d conformal anomaly \cite{Fuini:2015hba,Janiszewski:2015ura,Ghosh:2021naw}.  Again, since the tensor components in \eqref{DeltaTmunu} depend only on the magnetic field they do not contribute to the subtracted stress-energy tensor $\langle T^{\mu \nu} \rangle_r = \langle T^{\mu \nu} \rangle - \langle T^{\mu \nu} \rangle_{T=0}$ given by \eqref{EMT_rho_v2}-\eqref{EMT_P_z_v2}. The subtracted stress-energy tensor has a vanishing trace and is consistent with conformal magnetohydrodynamics.


\section{Numerical results for thermodynamic quantities}
\label{Sec:Numerics}

 In the previous section we obtained the holographic dictionary for the Gibbs free energy density and for the components of the stress-energy tensor of the 4d magnetised conformal plasma dual to the $AdS_5$ magnetic black brane. In this section we present our numerical results for several thermodynamic quantities. From the Gibbs free energy density we will derive  the entropy density, magnetisation, susceptibility and pyro-magnetic coefficient using the identities obtained in section \ref{Sec:CFT}. For the entropy density we will find agreement with the result obtained from the Bekenstein-Hawking area formula \eqref{entropy}. 
 
 In terms of the stress-energy tensor properties, we present our numerical results for the hydrodynamic pressures in the directions parallel and perpendicular to the magnetic field as well as the speeds of sound. We  show that these quantities display an evolution consistent with an RG flow of a 4d CFT in the regime of weak magnetic fields (high temperatures) to a 2d CFT in the regime of strong magnetic fields (low temperatures). 

We finish this section with a comparison of a number of thermodynamic quantities to the lattice QCD results found in \cite{Bali:2014kia}.

\subsection{Gibbs free energy and entropy densities}

\begin{figure}[ht!]
\centering
\includegraphics[width=0.4\textwidth]{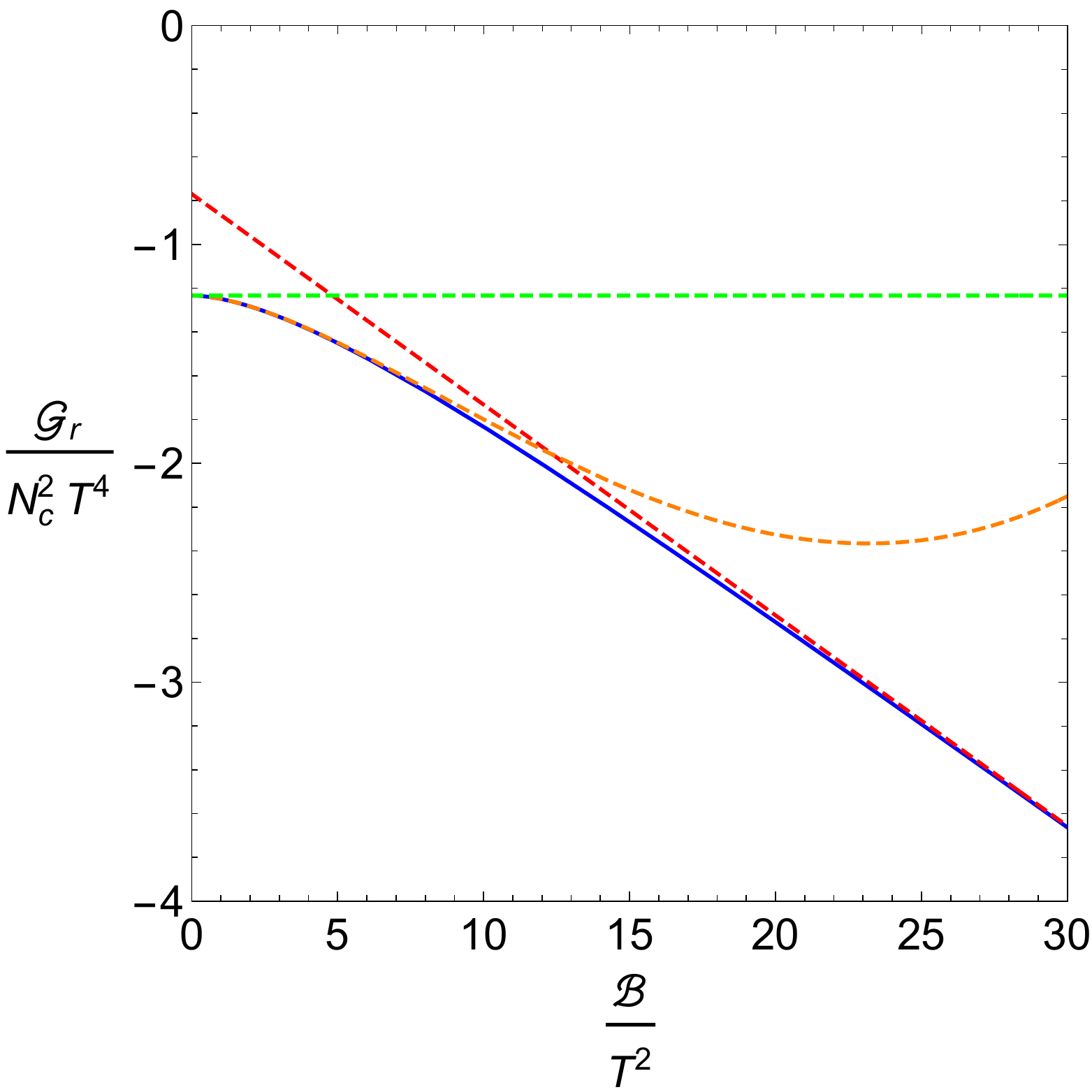}
\caption{The dimensionless free energy density $g_r={\cal G}_r/T^4$ (solid blue line), divided by $N_c^2$, as a function of $b={\cal B}/T^2$, compared with the analytical solution at small $b$ (orange dashed line) and large $b$ (red dashed line). The analytical solutions at small $b$ and large $b$, shown in equations \eqref{gRensmallb} and \eqref{gRenlargeb}, were obtained in appendices \ref{Appendix:smallb} and  \ref{Appendix:BTZ} respectively. The green dashed line depicts the  limit at $b=0$. }
\label{Fig:FreeEn}
\end{figure}

Using the results described in subsection \ref{subsubsec:Gibbs}  we numerically evaluate the dimensionless Gibbs free energy $g_r={\cal G}_r/T^4$, divided by $N_c^2$, as a function of the dimensionless ratio $b={\cal B}/T^2$. This is displayed as the  solid blue curve in figure \ref{Fig:FreeEn}. At small $b$ the dimensionless Gibbs free energy is well described by the formula:
\begin{equation}
g_r(b \ll 1) = N_c^2 \tilde \sigma \Big[ - \pi^4 + \frac13 b^2 
\left ( \tilde U_{\infty,4}(\sqrt{3}) + \ln (b/3 \pi^2) \right )  + \dots \Big ] \, ,  \label{gRensmallb}
\end{equation}
where  $\tilde U_{\infty,4}(\sqrt{3})\approx -0.25$, and  $\tilde \sigma = \sigma/N_c^2 = 1/(8 \pi^2)$ is the effective gravitational coupling, fixed previously to match the thermodynamics of the strongly coupled ${\cal N}=4$ super Yang-Mills plasma at ${\cal B}=0$. The analytic result in \eqref{gRensmallb} is derived in appendix \ref{Appendix:smallb} by considering a perturbative expansion around the solution with $b=0$. It is plotted as an orange dashed line in figure \ref{Fig:FreeEn}. 

At large $b$ the dimensionless Gibbs free energy is well described by the formula:
\begin{equation}
g_r(b \gg 1) = N_c^2 \Big [ - \frac{4}{3 \sqrt{3}} \pi^2 \tilde \sigma b + g_{\infty}  + \dots \Big ] \,   \label{gRenlargeb}
\end{equation}
where $g_{\infty}  \approx - 0.77$. This is displayed as a red dashed line in figure \ref{Fig:FreeEn}. The leading term in \eqref{gRenlargeb} is derived in appendix \ref{Appendix:BTZ} using the $\rm \bf BTZ \times \R^2$ solution and the sub-leading term is obtained from fitting to the full numerical solution. We would like to emphasise that the agreement between the thermodynamics of the $\rm \bf BTZ \times \R^2$ solution and the thermodynamics of the full numerical solution was possible thanks to the subtraction of the renormalised free energy with the zero temperature result described in the previous section. 

\begin{figure}[ht!]
\centering
\includegraphics[width=0.4\textwidth]{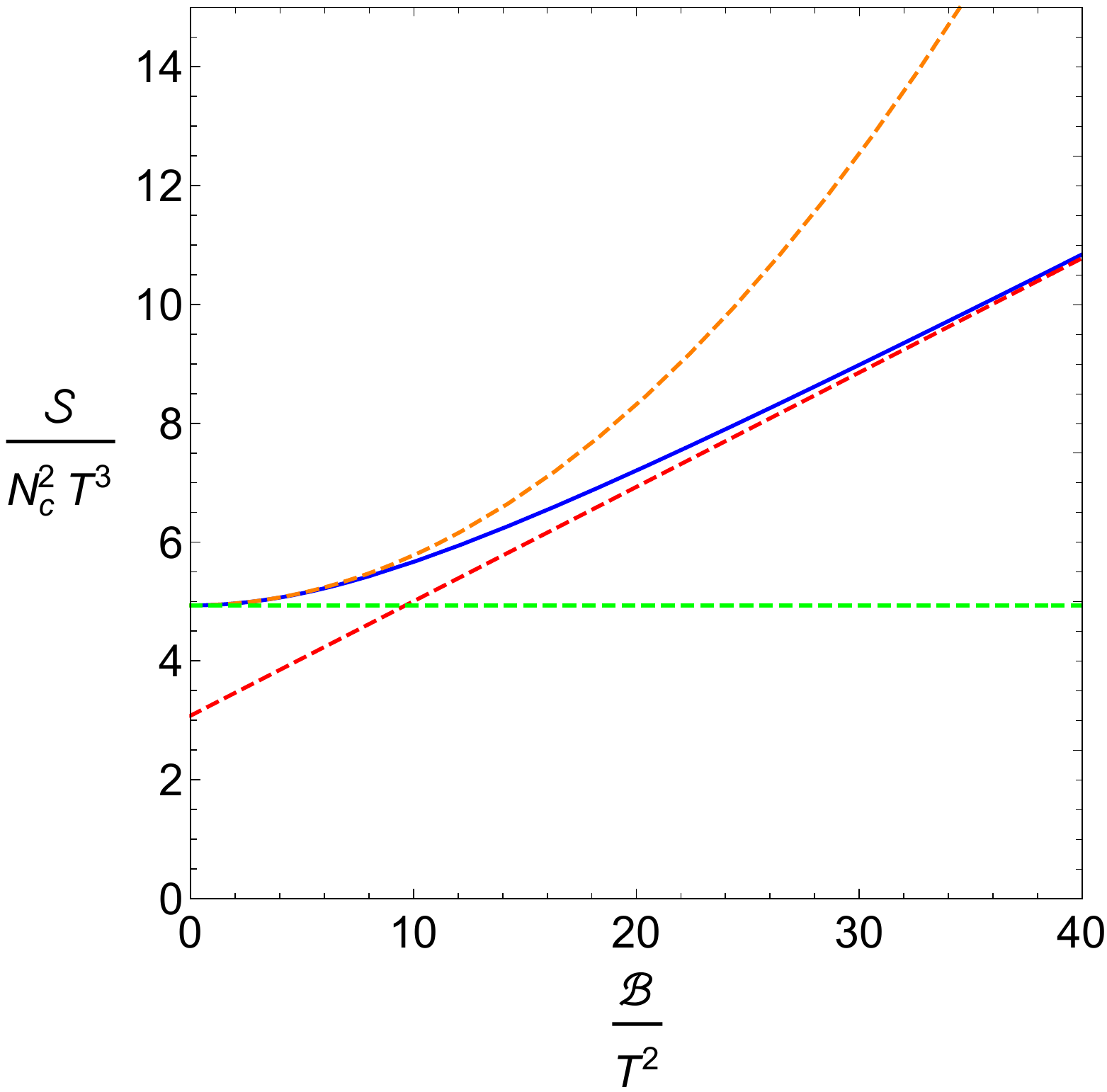}
\caption{The dimensionless entropy density $s={\cal S}/T^3$ (solid blue line), divided by $N_c^2$,  as a function of $b={\cal B}/T^2$, compared with the analytical solution at small $b$ (orange dashed line) and large $b$ (red dashed line). The analytical solutions at small $b$ and large $b$, shown in equations \eqref{entsmallb} and \eqref{entlargeb}, were obtained in appendices \ref{Appendix:smallb} and  \ref{Appendix:BTZ} respectively. The green dashed line depicts the Stefan-Boltzmann limit at $b=0$.}
\label{Fig:Entropy}
\end{figure}

The dimensionless entropy density is obtained from the thermodynamic relation 
\begin{equation}
s = \frac{{\cal S}}{T^3}  
=  2 b \, g_r'(b) - 4  g_r(b)\, , \label{sfromg}
\end{equation}
found in section \ref{Sec:CFT} for a 4d conformal magnetic fluid. Note that the quadratic term in \eqref{grfromtG} does not contribute to the dimensionless entropy density. In figure \ref{Fig:Entropy} we plot the dimensionless entropy density $s={\cal S}/T^3$, divided by $N_c^2$, as a function of $b={\cal B}/T^2$. The solid blue line represents the full numerical result whereas the orange and red dashed lines correspond to the analytical results at small and large $b$ respectively, found in appendices \ref{Appendix:smallb} and \ref{Appendix:BTZ}. The green dashed line represents the limit $s = 4 \pi^4 N_c^2 \bar \sigma = (\pi^2/2) N_c^2$ for the dimensionless entropy at $b=0$.

Our results for the entropy density ${\cal S}$, obtained from the renormalised Gibbs free energy ${\cal G}_r$, are equivalent to those obtained from the Bekenstein-Hawking formula \eqref{entropy} and are compatible with those obtained in \cite{DHoker:2009mmn}. This is a non-trivial consistency check for our holographic renormalisation procedure.  Notice that the formula \eqref{sfromg} is valid in general as long as we consider the subtracted Gibbs free energy density ${\cal G}_r = {\cal G}(T) - {\cal G}(0)$ which is consistent with conformal invariance as described in the previous section.

The subtraction was also important in order to reproduce the result for the Gibbs free energy density associated with the $\rm \bf BTZ \times \R^2$ solution at large $b$ from the full numerical solution, as described in the previous subsection. 

At small $b$ the entropy density is well described by the formula:
\begin{equation}
s (b \ll 1) =  N_c^2 \tilde \sigma \Big [ 4 \pi^4 + \frac23 b^2 + \dots \Big ]   \, , \label{entsmallb}
\end{equation}
found using equations \eqref{gRensmallb} and \eqref{sfromg}.
At large $b$, the entropy density takes the BTZ form
\begin{equation}
s (b \gg  1) =  N_c^2 \Big [ \frac{8}{3 \sqrt{3}}  \pi^2 \tilde \sigma b - 4 g_{\infty} + \dots \Big ] \, , \label{entlargeb}
\end{equation}
found using \eqref{gRenlargeb} and \eqref{sfromg}.

\subsection{Magnetisation, susceptibility and the pyro-magnetic coefficient}
\label{subsec:magnetisation}

\begin{figure}[ht!]
\centering
\includegraphics[width=0.4\textwidth]{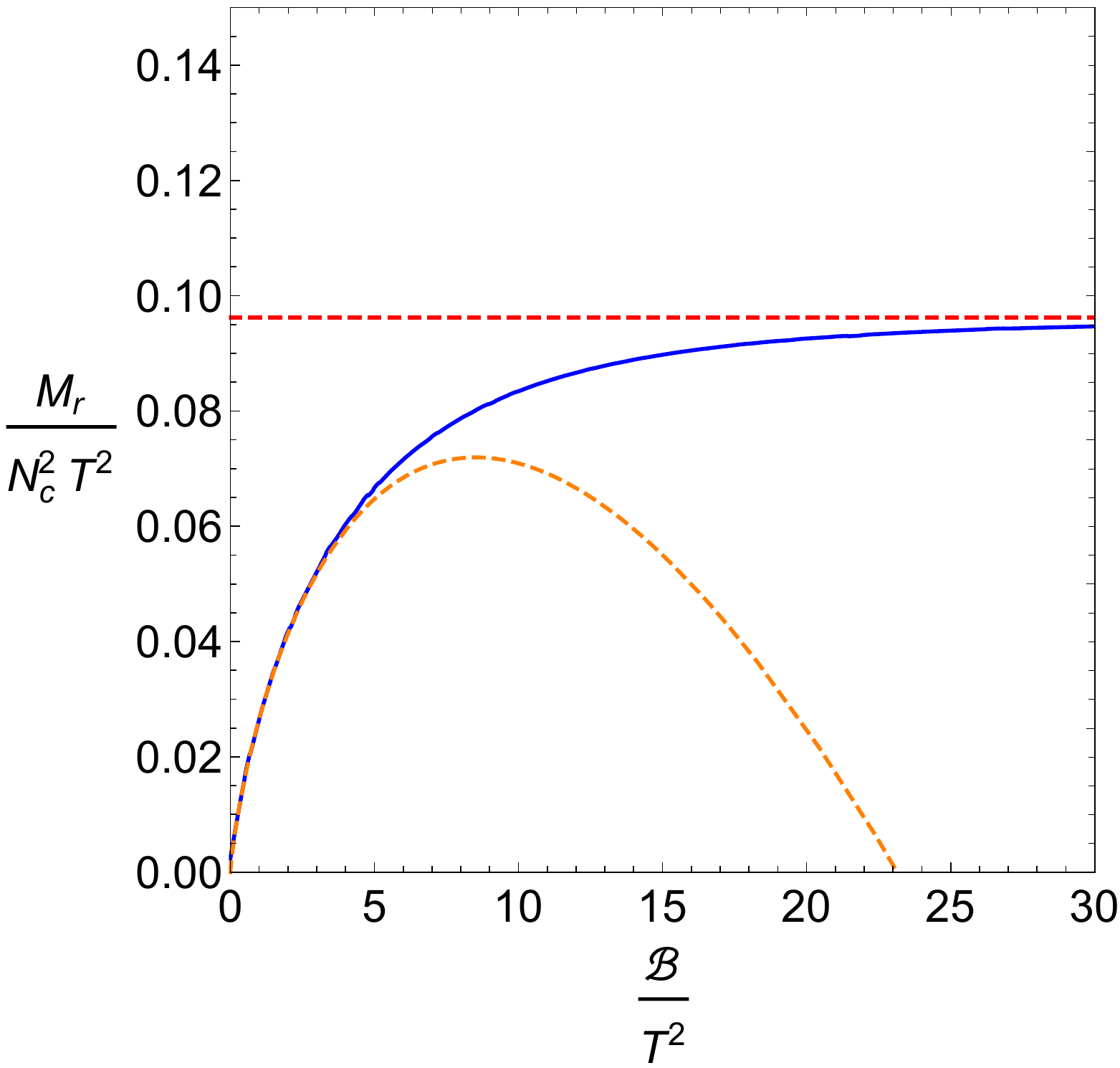}
\caption{The dimensionless magnetisation  $m_r= M_r/T^2$, divided by $N_c^2$, as a function of $b={\cal B}/T^2$ (solid blue  line), compared with the analytical solutions at small $b$ (orange dashed line) and large $b$ (red dashed line).   The analytical solutions at small $b$ and large $b$ are derived in appendices \ref{Appendix:smallb} and \ref{Appendix:BTZ}.}
\label{Fig:mRen}
\end{figure}

We calculate the dimensionless magnetisation density using the formula
\begin{equation}
m_r = \frac{ M_r}{T^2} = - g_r'(b) \, ,  \label{mfromg}  
\end{equation}
obtained in section \ref{Sec:CFT}. Our results for the dimensionless magnetisation $m_r$, divided by $N_c^2$, as a function of $b={\cal B}/T^2$ are displayed in figure \ref{Fig:mRen}. The solid blue line represents the full numerical result whereas the orange and red dashed lines correspond to the analytical results at small and large $b$ respectively, found in appendices \ref{Appendix:smallb} and \ref{Appendix:BTZ}. 

We observe that the dimensionless magnetisation density $m_r=M_r/T^2$ increases monotonically from zero at $b=0$ to a positive constant value in the limit  $b \to \infty$. The latter corresponds to the regime of very high values of ${\cal B}$ at fixed $T$ or very low values of $T$ at fixed ${\cal B}$. In this limit we expect a dimensionality reduction from a 4d CFT to a 2d CFT. Indeed, the asymptotic form of the magnetisation density coincides with the analytical result found in appendix \ref{Appendix:BTZ} for the  $\rm \bf BTZ \times \R^2$ solution.

\begin{figure}[ht!]
\centering
\includegraphics[width=0.4\textwidth]{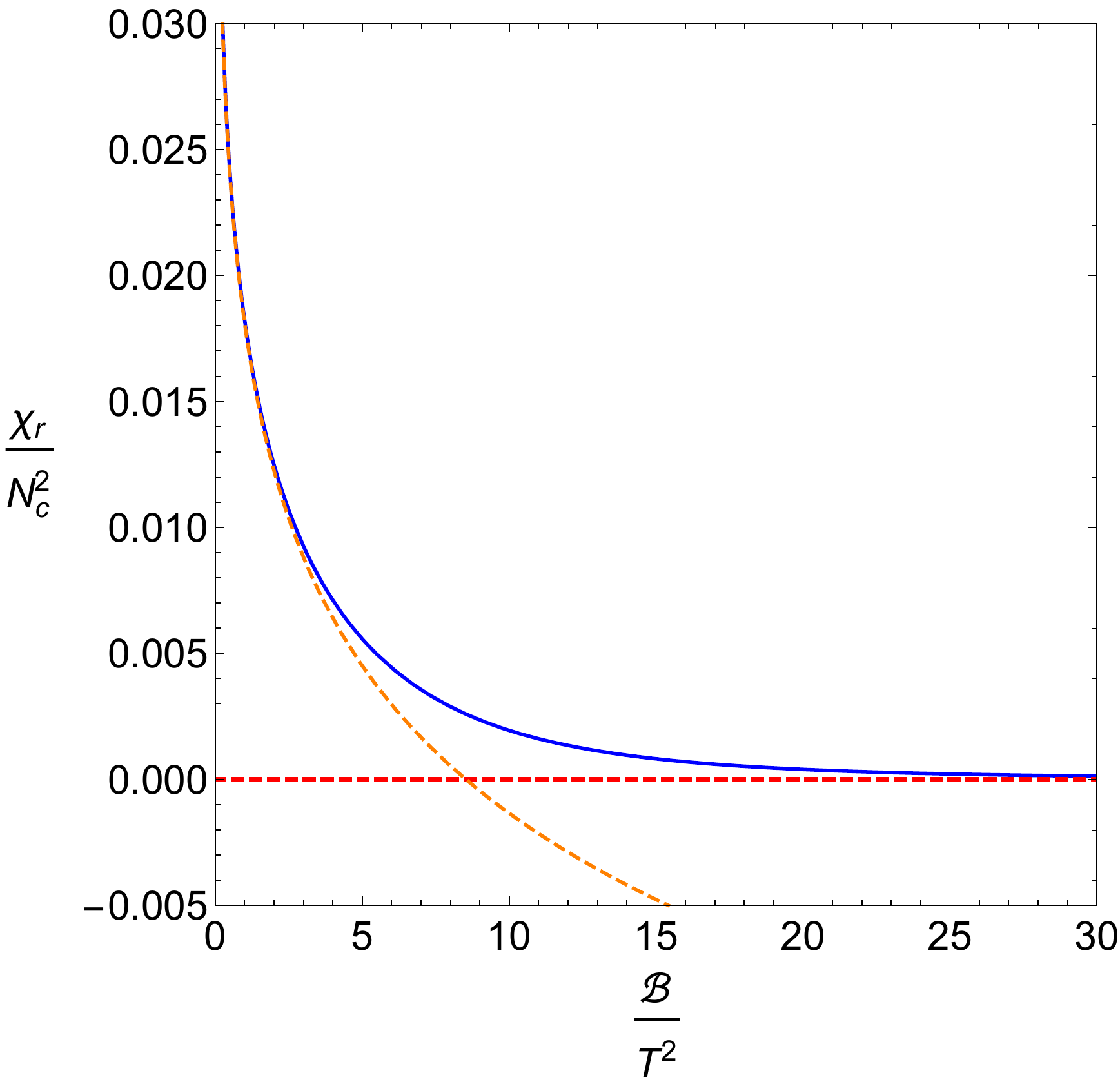}
\hspace{1cm}
\includegraphics[width=0.38\textwidth]{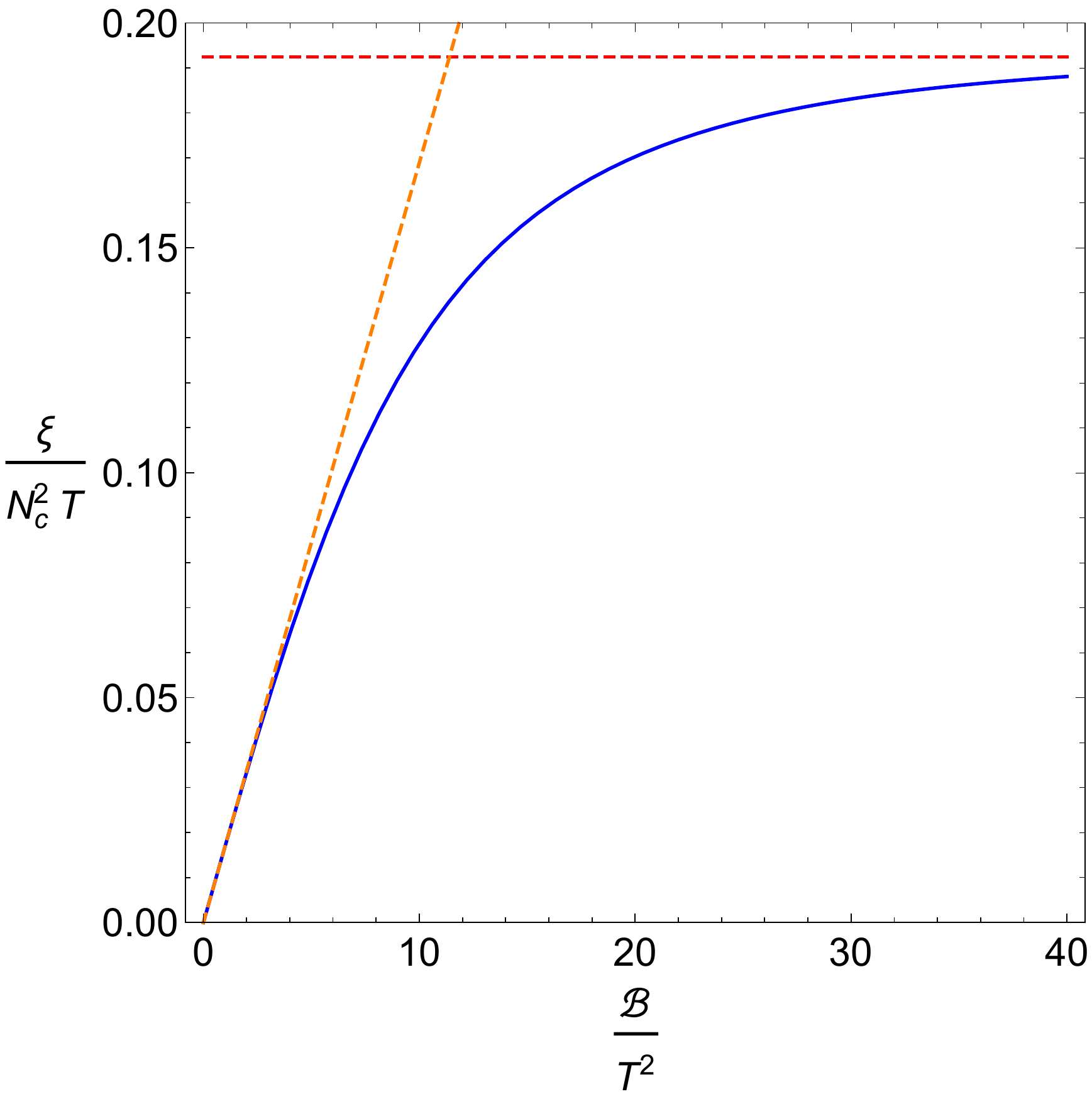}
\caption{Left panel: magnetic susceptibility $\chi_r$, divided by $N_c^2$,  as a function of $b={\cal B}/T^2$. Right panel: Dimensionless pyro-magnetic coefficient $\xi/T$, divided by $N_c^2$, as a function of $b={\cal B}/T^2$. In both panels the blue lines represent our numerical results whilst the orange and red dashed lines represent the analytical solutions at small and large $b$ respectively, obtained in appendices \ref{Appendix:smallb} and \ref{Appendix:BTZ}. }
\label{Fig:chiRen}
\end{figure}

We also calculate the magnetic susceptibility $\chi$ and the dimensionless pyro-magnetic coefficient $\xi/T$  using the formulae 
\begin{align}
\chi = - g_r''(b) \quad , \quad 
\frac{\xi}{T} = 2 \Big [ b g_r''(b) - g_r'(b) \Big ]    \, ,  \label{chixifromg}
\end{align}
found in section \ref{Sec:CFT}. Our results for the magnetic susceptibility and the dimensionless pyro-magnetic coefficient, both divided by $N_c^2$, are displayed on the left and right panels of figure \ref{Fig:chiRen} respectively. In both plots the blue solid curve represents the full numerical result whereas the orange and red dashed lines represent the analytical results at small and large $b$, found in appendices \ref{Appendix:smallb} and \ref{Appendix:BTZ}.  

As shown in figure \ref{Fig:chiRen}, the magnetic susceptibility $\chi$ is a non-negative monotonically decreasing function of $b$ whilst the dimensionless pyro-magnetic coefficient $\xi/T$ is a non-negative monotonically increasing function of $b$ . Note that $\chi$ diverges in the limit $b \to 0$ which corresponds to very low values of ${\cal B}$ at fixed $T$ or very high values of $T$ at fixed ${\cal B}$.  
This divergence is logarithmic, as can be seen from \eqref{gRensmallb} and \eqref{chixifromg}. In the opposite limit, $b \to \infty$, the magnetic susceptibility vanishes. Meanwhilst, the dimensionless pyro-magnetic coefficient vanishes in the limit $b \to 0$ and reaches a positive constant value in the limit $b \to \infty$.

At the end of this section we will compare the results for the magnetisation density and the susceptibility in the ${\cal N}=4$ Super Yang-Mills plasma against lattice QCD results. 

\subsection{The anisotropic pressures}

The anisotropy of the 4d conformal fluid due to the presence of the magnetic field is characterised by the difference between the pressure along the direction parallel to the magnetic field ($P_z$) and the pressure along the directions transverse to the magnetic field ($P_x=P_y$).  The holographic dictionary for the longitudinal and transverse pressures $P_z$ and $P_x$ was obtained in subsection \ref{subsec:Hologstress} in terms of the UV coefficients of the metric. As an alternative method, we have also evaluated the hydrodynamic pressures using the identities $P_z = - {\cal G}$ and $P_x=P-M {\cal B}$ found at the end of section \ref{Sec:CFT}. Both methods yield the same results, which is a non-trivial consistency check of our holographic renormalisation procedure. In appendix \ref{Appendix:smallb}, where we used an analytical solution at small $b$, the agreement between the pressures obtained from the stress tensor holographic dictionary and the results using the identities $P_z = - {\cal G}$ and $P_x=P-M {\cal B}$
is explicit. 

\begin{figure}[ht!]
    \centering
    \includegraphics[width=0.44\textwidth]{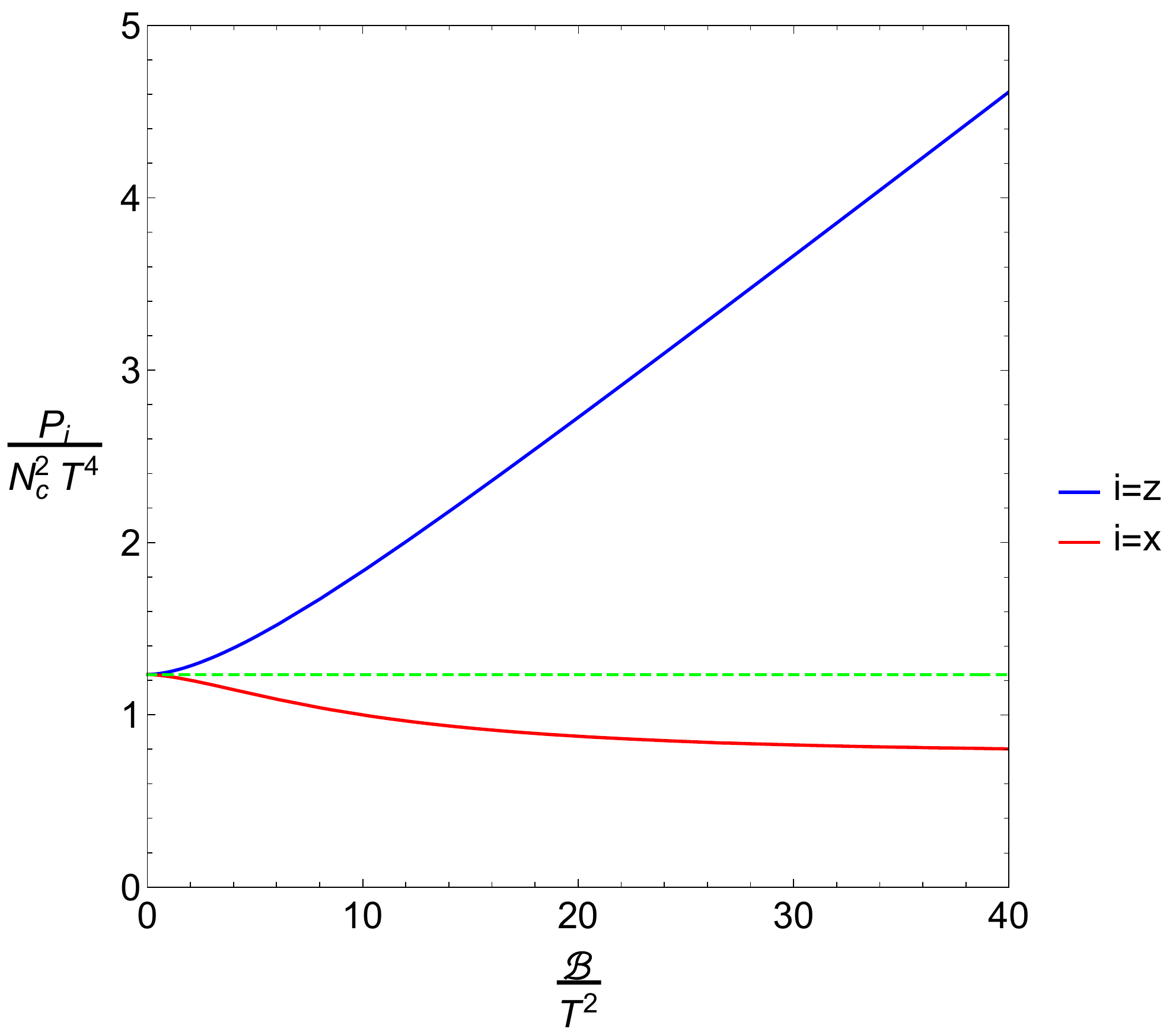}
    \caption{The dimensionless longitudinal pressure $P_z/T^4$ (blue) and the dimensionless transverse pressure $P_x/T^4$ (red), both divided by $N_c^2$, as functions of $b={\cal B}/T^2$. The green dashed line depicts the limiting case $b=0$.}
    \label{fig:RenPressures}
\end{figure}

In figure \ref{fig:RenPressures} we present our numerical results for the hydrodynamic pressures. The solid blue  line represents the evolution of the dimensionless longitudinal pressure $P_z/T^4$, divided by $N_c^2$, with $b={\cal B}/T^2$ whereas the solid red line represents the evolution of the dimensionless transverse pressure $P_x/T^4$, divided by $N_c^2$, with  $b={\cal B}/T^2$.  The green dashed horizontal line corresponds to the limiting case $b=0$ where isotropy is restored in the conformal fluid. From figure \ref{fig:RenPressures} it is clear that the difference between the longitudinal and transverse pressures increases with ${\cal B}/T^2$ which means that the anisotropy increases when the magnetic field increases or the temperature decreases. 
At the end of this section we will compare the results for the anisotropic pressures in a magnetised conformal plasma against the  results found in lattice QCD for the magnetised quark-gluon plasma. 

\subsection{Specific heats and speeds of sound}

The specific heat is obtained from the thermodynamic relation 
\begin{equation}
C_{V,{\cal B}} = T^3 \Big [ -12 g_r(b) + 10 b g_r'(b) - 4 b^2 g_r''(b) \Big ]  \,, \label{cV}
\end{equation}
obtained in section \ref{Sec:CFT}. Since the specific heat is related to the entropy density, the quadratic term in the renormalised Gibbs free energy density  \eqref{grfromtG} does not contribute to it. 
In section \ref{Sec:CFT} we also showed that the squared speeds of sound along the $z$ and $x$ directions are given by
\begin{equation}
c_{s,x}^2 =   \frac{ {\cal S} - \xi {\cal B}}{C_{V,{\cal B}}}   \quad , \quad 
c_{s,z}^2 =  \frac{{\cal S}}{C_{V,{\cal B}}} \,,
\end{equation} 
where ${\cal S}$ is the entropy density, $\xi$ the pyromagnetic coefficient and $C_{V,{\cal B}}$ the specific heat.

\begin{figure}[ht!]
    \centering
    \includegraphics[width=0.4\textwidth]{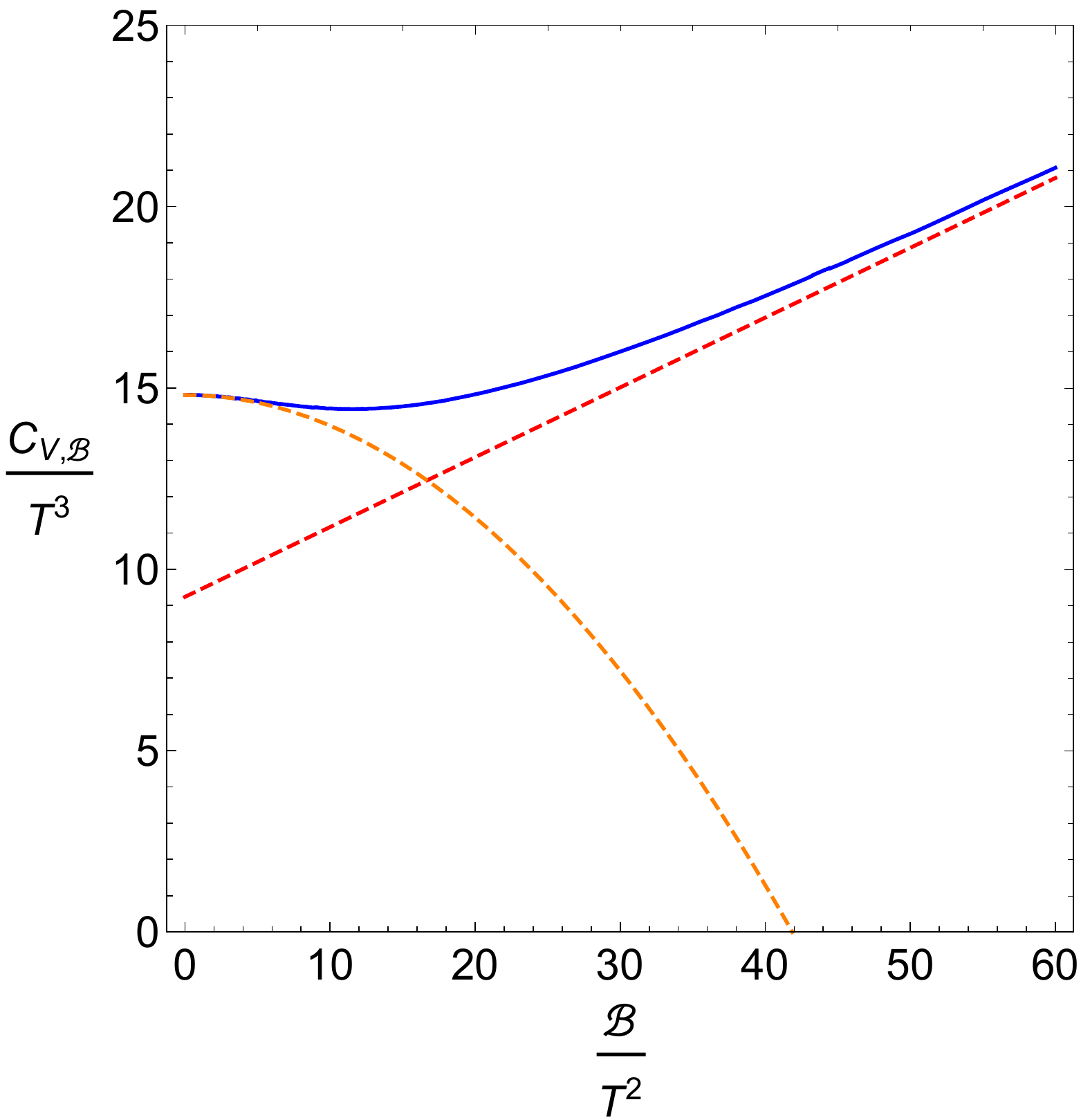}
            \caption{Dimensionless specific heat $C_{V,{\cal B}}/T^3$ (solid)  as a function of $b={\cal B}/T^2$ (solid blue  line), compared with the analytical solutions at small $b$ (orange dashed line) and large $b$ (red dashed line).   The analytical solutions at small $b$ and large $b$ were obtained in appendices \ref{Appendix:smallb} and \ref{Appendix:BTZ}.  }
    \label{fig:cB}
\end{figure}

Figure \ref{fig:cB} shows our numerical results for the dimensionless specific heat $C_{V,{\cal B}}/T^3$ as a function of $b={\cal B}/T^2$ (solid blue line) compared with the analytical solutions at small $b$ (orange dashed line) and large $b$ (red dashed line). The latter are obtained by plugging \eqref{gRensmallb} and \eqref{gRenlargeb} into \eqref{cV}. We observe that $C_{V,{\cal B}}/T^3$ is a non-monotonic function of $b={\cal B}/T^2$ and reaches a linear asymptotic behaviour consistent with the $\rm \bf BTZ \times \R^2$ analytic solution only at very large values of $b$.

\begin{figure}[ht!]
    \centering
    \includegraphics[width=0.44\textwidth]{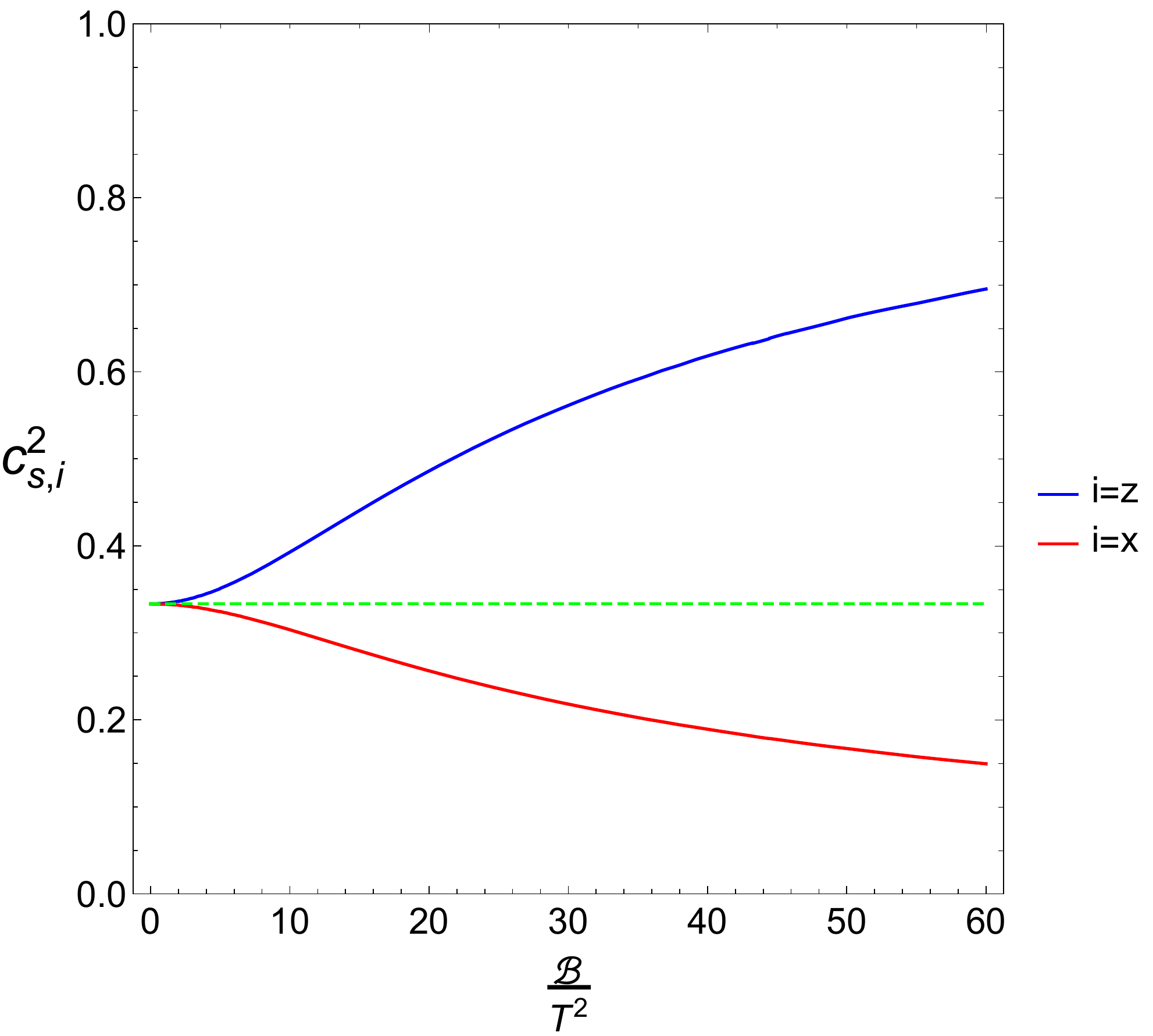}
    \hspace{1cm}
    \includegraphics[width=0.4\textwidth]{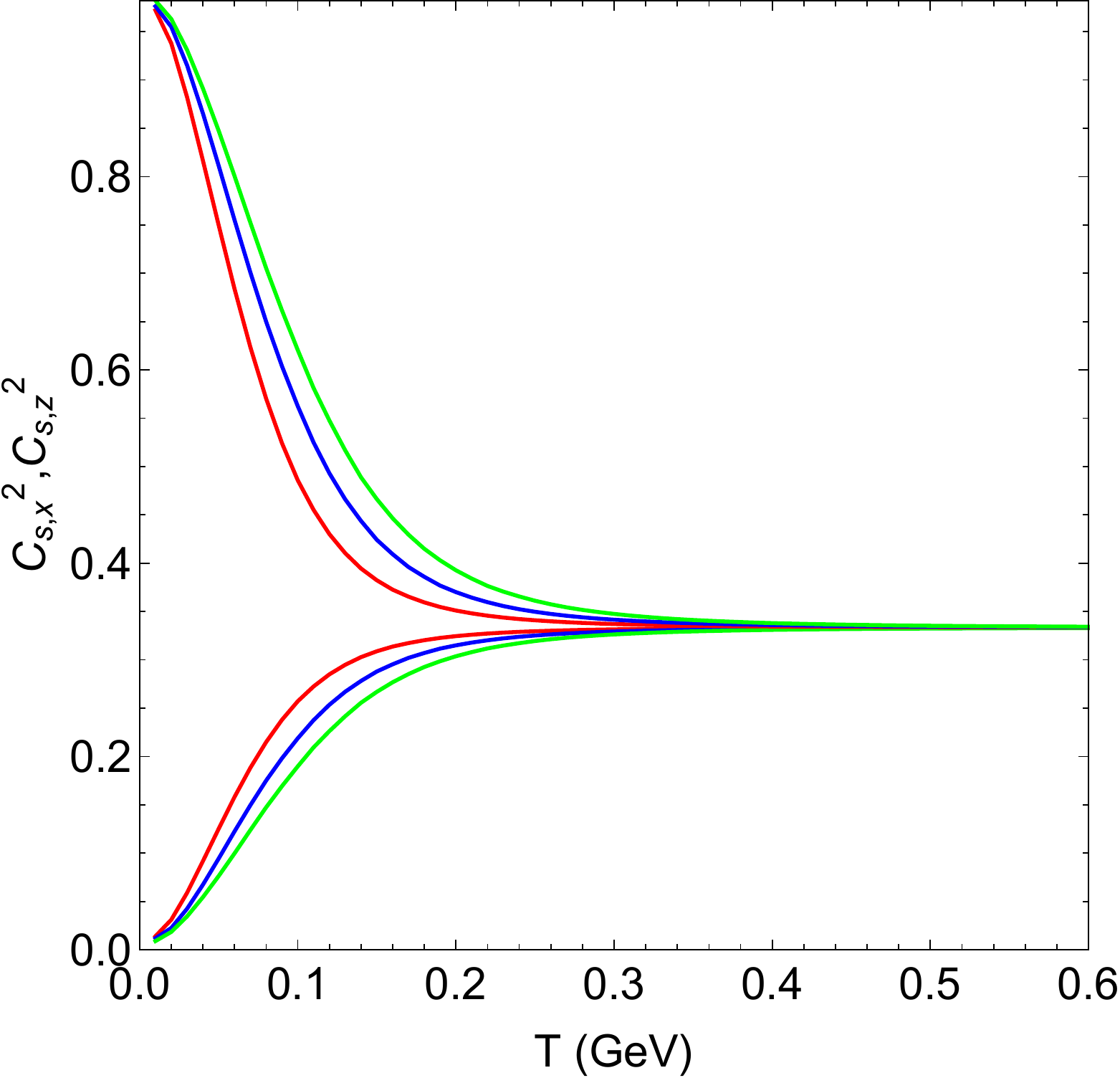}
        \caption{{\bf Left panel:} Squared speeds of sound  along the $z$ and $x$ directions (blue and red respectively) as a function of $b={\cal B}/T^2$. The green dashed horizontal line represents the isotropic result $c_{s,x}^2=c_{s,z}^2=1/3$ in the absence of magnetic field. {\bf Right panel:} Squared speeds of sound as a function of $T$ for fixed values of ${\cal B}$. The red, blue and green lines correspond to ${\cal B}=0.2 \, \text{GeV}^2$, ${\cal B}=0.3 \, \text{GeV}^2$ and ${\cal B}=0.4 \, \text{GeV}^2$ respectively. The asymptotic values of $c_{s,z}^2$ and $c_{s,x}^2$ in the large $b$ limit are $1$ and $0$ respectively. This corresponds to the $T \to 0$ limit on the right panel. }
    \label{fig:cs2}
\end{figure}

Our results for the speeds of sound along the $z$ direction (parallel to the magnetic field) and $x$ direction (transverse to the magnetic field) are presented in figure \ref{fig:cs2}. The plot on the left panel displays the squared speeds of sound as a function of $b={\cal B}/T^2$. The blue and red lines correspond to $c_{s,z}^2$ and $c_{s,x}^2$ respectively. The green dashed line represents the limiting case $b=0$ where $c_{s,z}^2=c_{s,x}^2=1/3$. The plot on the right panel displays the squared speeds of sound as functions of the temperature $T$ for fixed values of the magnetic field ${\cal B}$. The red, blue and green lines correspond to ${\cal B}=0.2 \, \text{GeV}^2$, ${\cal B}=0.3 \, \text{GeV}^2$ and ${\cal B}=0.4 \, \text{GeV}^2$ respectively. Note that as the temperature decreases the squared speed of sound $c_{s,z}^2$ grows from $1/3$ to $1$ whereas the squared speed of sound  $c_{s,x}^2$ decreases from $1/3$ to $0$. This is consistent with an RG flow of the 4d CFT at high temperatures to a 2d CFT low temperatures. The same pattern occurs when increasing the magnetic field ${\cal B}$ for fixed values of $T$.  At small $b$ the squared speeds of sound are well described by the analytic expressions \eqref{cs2zsmallb} and \eqref{cs2xsmallb} of appendix \ref{Appendix:smallb} whilst the large $b$ limit $c_{s,z}^2 \to 1$ for the longitudinal direction and $c_{s,x}^2 \to 0$ 
for the transverse direction is consistent with the $\rm \bf BTZ \times \R^2$ solution, as confirmed in \eqref{cs2xzlargeb} of appendix \ref{Appendix:BTZ}.

\begin{figure}[ht!]
    \centering
    \includegraphics[width=0.4\textwidth]{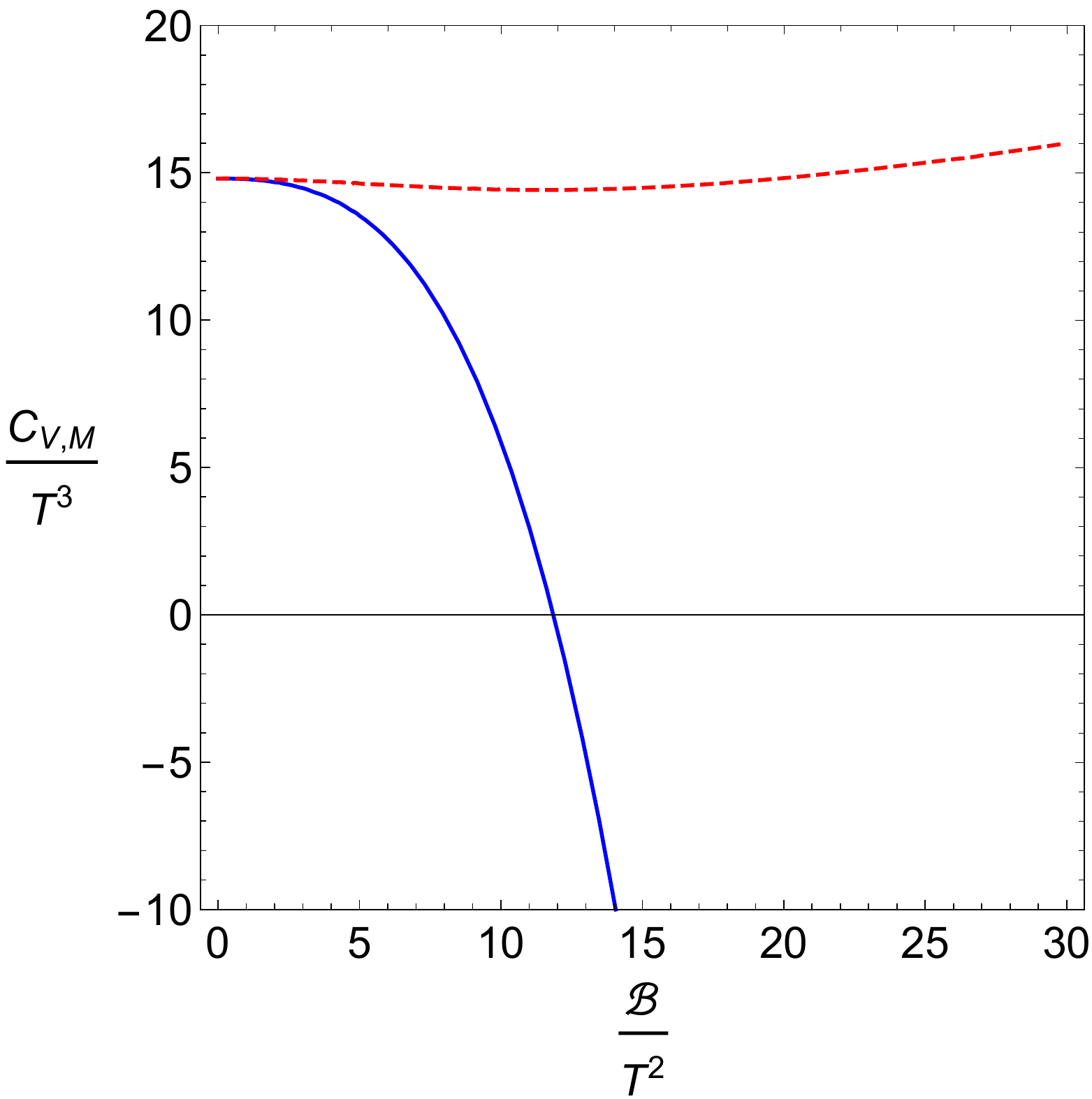}
            \caption{Dimensionless specific heat at fixed magnetisation $C_{V,M}/T^3$   as a function of $b={\cal B}/T^2$ (blue solid line), compared against the dimensionless specific heat at fixed magnetic field $C_{V,{\cal B}}/T^3$ (red dashed line).  }
    \label{fig:cM}
\end{figure}

We finish this subsection with a brief analysis of local thermal equilibrium, based on \cite{Landau:1980mil}. The conditions for local thermal equilibrium follow from considering the variation of the Gibbs free energy $G=U-TS-M {\cal B}$ due to small variations of the extensive quantities $S$ and ${\cal M}$.\footnote{The quantities $M$ and ${\cal B}$ here correspond to $V$ and $-P$ in \cite{Landau:1980mil}.} Expanding the internal energy $U$ in terms of $S$ and $M$ and imposing that the variation of $G$ is positive (condition of minimum at equilibrium) one finds the following two conditions:
\begin{equation}
\left ( \frac{ \partial T}{ \partial S} \right )_M  = \frac{T}{C_{V,M}}  >0   \quad , \quad
\left ( \frac{\partial {\cal B}}{ \partial M} \right )_T
= \frac{1}{\chi} >0 \, ,
\end{equation}
where $C_{V,M}$ is the specific heat at constant magnetisation and $\chi$ the magnetic susceptibility. We note that the second condition is automatically satisfied  in our framework because the magnetic susceptibility found in this work is always positive, as shown in figure \ref{Fig:chiRen}. As regards the first condition, following \cite{Landau:1980mil} we obtain a relation between $C_{V,M}$ (specific heat at fixed magnetisation) and $C_{V,{\cal B}}$ (specific heat at fixed magnetic field):
\begin{eqnarray}
C_{V,M} = C_{V,{\cal B}} - T \frac{ ( \partial M / \partial T)_{\cal B}^2 } {(\partial M / \partial {\cal B})_T} 
= C_{V, {\cal B}} - T \frac{\xi^2}{\chi} \,, 
\end{eqnarray}
where $\xi$ is the pyro-magnetic coefficient and $\chi$ the magnetic susceptibility, both described in subsection \ref{subsec:magnetisation}. Using this formula we can evaluate the specific heat $C_{V,M}$ in terms of the thermodynamic quantities found before. Figure \ref{fig:cM} shows the dimensionless specific heat $C_{V,M}/T^3$   as a function of $b={\cal B}/T^2$ (blue solid line), compared against the dimensionless specific heat $C_{V,{\cal B}}/T^3$ (red dashed line). Interestingly, this plot indicates the emergence of an instability at around ${\cal B}=12 T^2$. Similar instabilities have been found previously in anisotropic backgrounds \cite{Mateos:2011ix,Mateos:2011tv}.

\subsection{Comparison against lattice QCD}

We end this section with a phenomenological analysis
where we compare some of the thermodynamic quantities of a  (strongly coupled) magnetised conformal plasma to lattice QCD results for the quark-gluon plasma \cite{Bali:2014kia}. 

For this comparison, we depart from the strongly coupled ${\cal N}=4$ Super Yang-Mills plasma, where the 5d gravitational constant $\sigma = 1/(16 \pi G_5) = N_c^2/(8 \pi^2)$ was fixed by string theory. We will instead consider  a phenomenological approach where $\sigma$ is fixed in such a way that in the absence of the magnetic field we reproduce the result for the free energy density of a free Yang-Mills plasma in the large $N_c$ limit. Namely, we will take $\sigma = N_c^2/(45 \pi^2)$ so that we recover the Stefan-Boltzmann result 
\begin{equation}
{\cal G}_{{\cal B}=0} = - \frac{\pi^2}{45} N_c^2 T^4 \,.  
\end{equation}
 This phenomenological approach was successfully pursued in the improved holographic QCD models of \cite{Gursoy:2008bu,Gursoy:2009jd}.

\begin{figure}[ht!]
\centering
\includegraphics[width=0.44\textwidth]{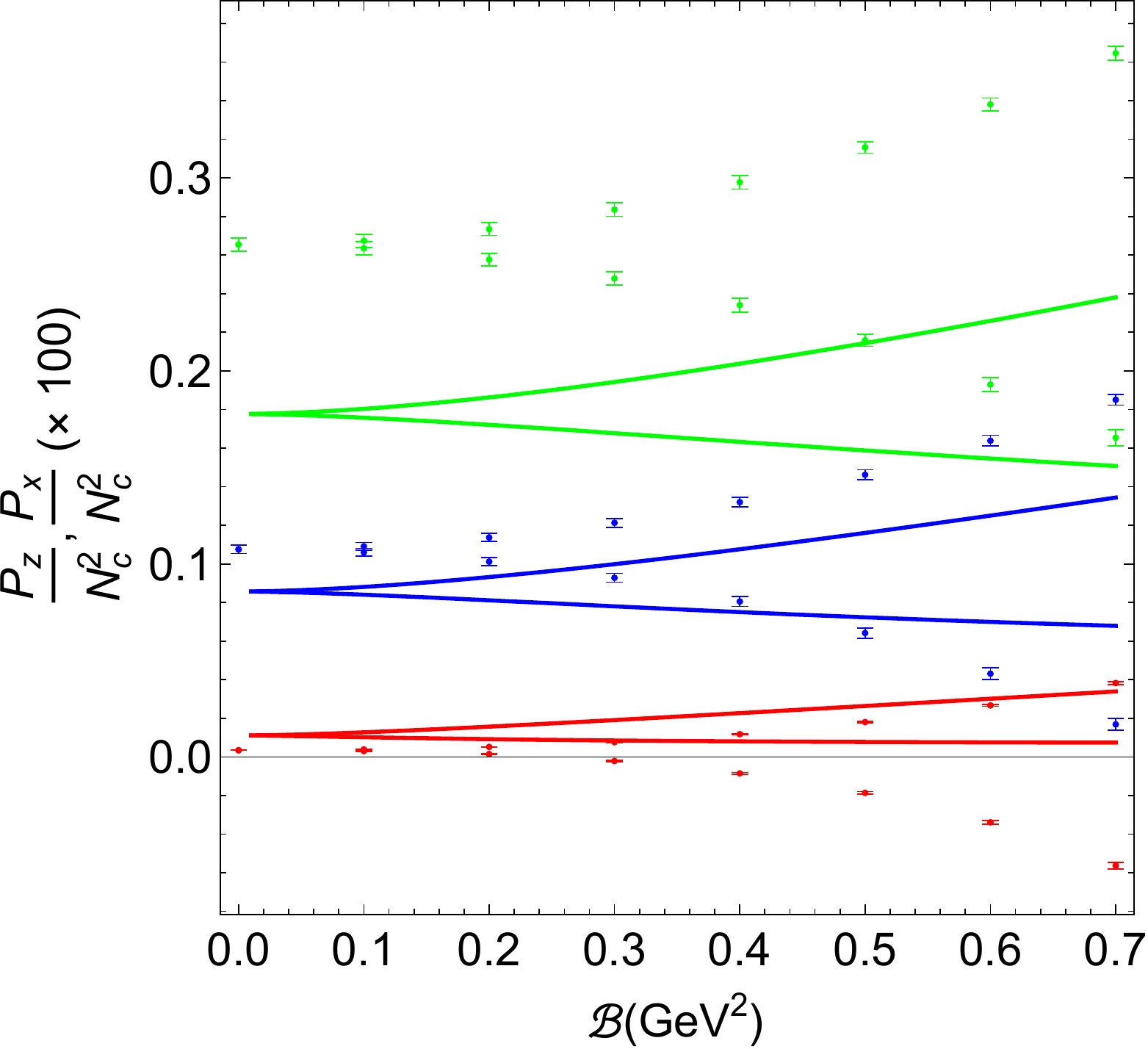}
\caption{Longitudinal pressure $P_z$ (upper solid line for a given colour) and transverse pressure $P_x$ (lower solid line for a given colour), both divided by $N_c^2$, as functions of the magnetic field ${\cal B}$ for fixed values of the temperature, compared with lattice results (dots with error bars), divided by $N_c^2=3^2$, obtained in \cite{Bali:2014kia}. The red, blue and green colours correspond to  $T=0.15 \,{\rm GeV}$, $0.25 \, {\rm GeV}$ and $0.3 \, {\rm GeV}$ respectively. }
\label{Fig:PressureLattice}
\end{figure}

We start this phenomenological analysis by comparing our results for the longitudinal and transverse pressures of a magnetised conformal plasma  against the lattice QCD results for the pressures of a magnetised quark-gluon plasma \cite{Bali:2014kia}, as shown in figure \ref{Fig:PressureLattice}. The figure shows the evolution of the pressures, divided by $N_c^2$, with the magnetic field ${\cal B}$ for fixed values of the temperature. The red, blue and green colours correspond to  $T=0.15 \,{\rm GeV}$, $0.25 \, {\rm GeV}$ and $0.3 \, {\rm GeV}$ respectively. For each colour, the upper and lower solid lines represent our results for the longitudinal and transverse pressures, divided by $N_c^2$, respectively. Likewise, the upper and lower dots with error bars represent the lattice QCD results for the longitudinal and transverse pressures (divided by $N_c^2=3^2$). The main conclusion from this figure is that the anisotropy between the longitudinal pressure $P_z$ and the transverse pressure $P_x$ in a  magnetised conformal plasma increases with the magnetic field in a way that is qualitatively similar to the anisotropic increase found for the QCD quark-gluon plasma. 

Note that at small temperatures (red) the holographic model does not reproduce quantitatively the lattice QCD anisotropy (the angle between the lines corresponding to $P_z$ and $P_x$ in the figure). This is because confinement and chiral symmetry breaking effects are important in this region.  At high temperatures (blue and green) the holographic model captures well the anisotropy increase between the longitudinal and transverse pressures.

\begin{figure}[ht!]
\centering
\includegraphics[width=0.43\textwidth]{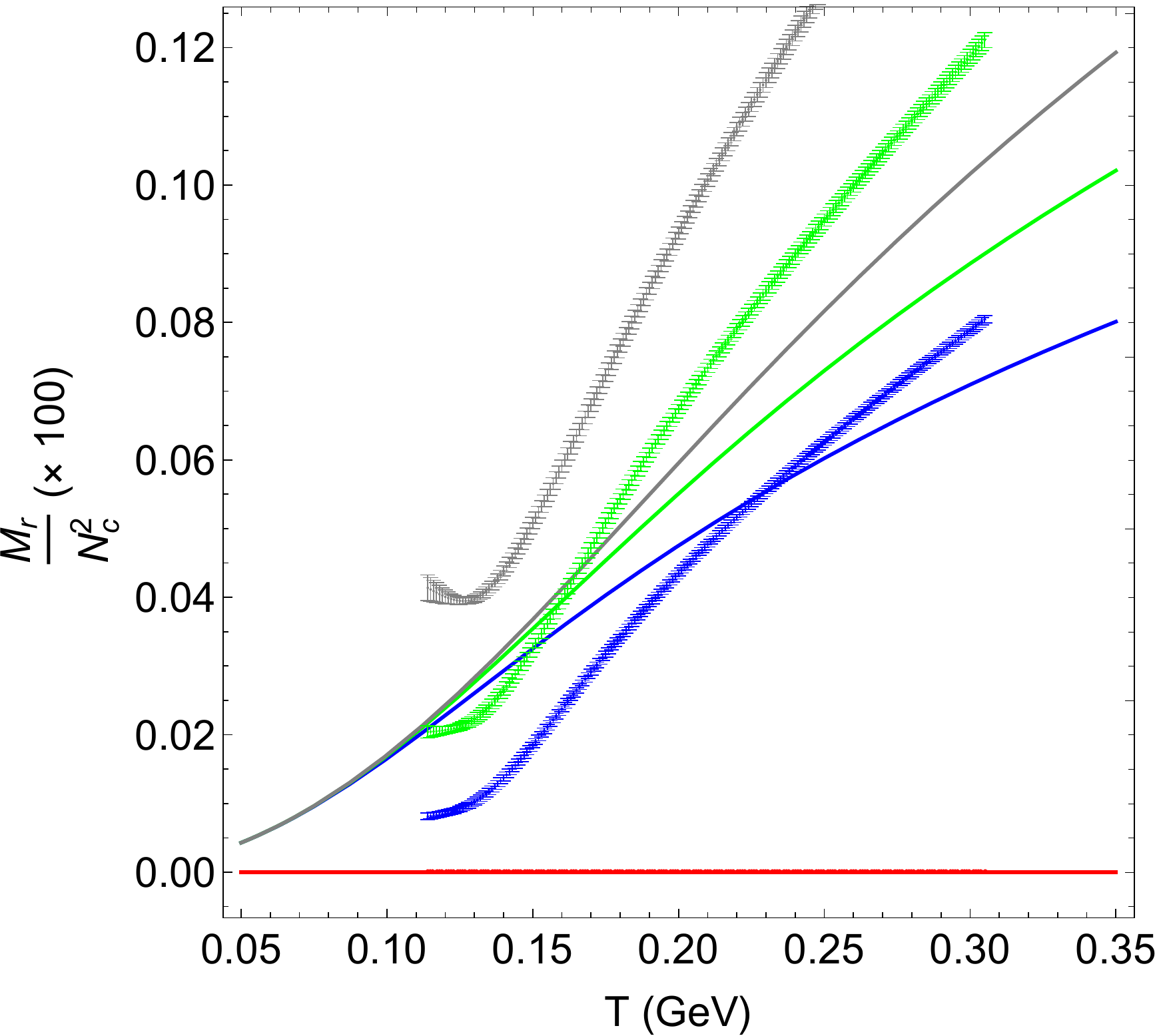}
\hspace{1cm}
\includegraphics[width=0.425\textwidth]{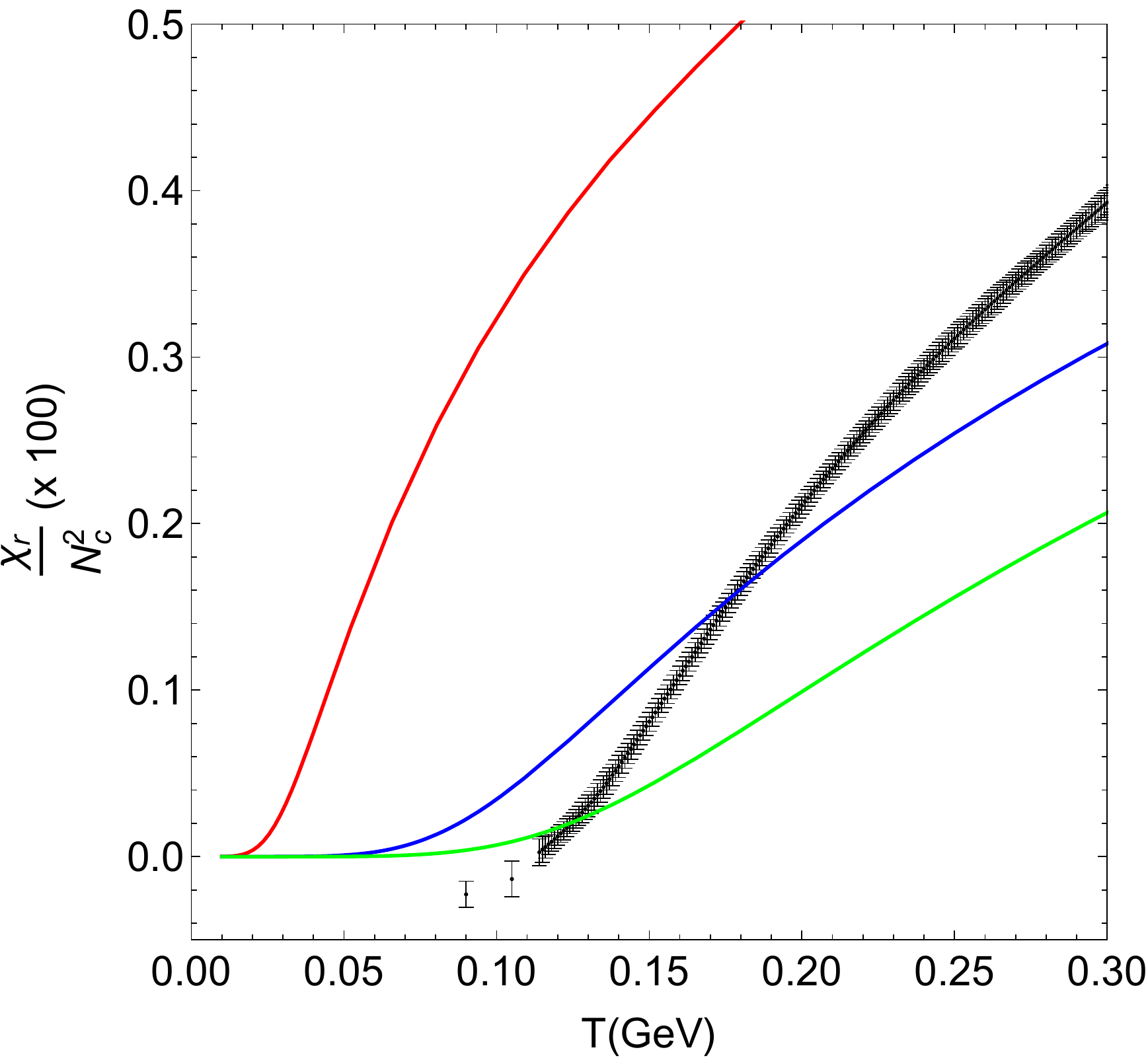}
\caption{{\bf Left panel:} Magnetisation density $M_r$, divided by $N_c^2$, as a function of the temperature (solid lines), compared with lattice results (dots with error bars), divided by $N_c^2=3^2$, for fixed values of the magnetic field. The red, blue, green and gray lines correspond to ${\cal B}=0 \, {\rm GeV}^2$, ${\cal B}=0.2 \, {\rm GeV}^2$, ${\cal B}=0.3 \, {\rm GeV}^2$ and ${\cal B}=0.4 \, {\rm GeV}^2$. {\bf Right panel:} Magnetic susceptibility $\chi_r$, divided by $N_c^2$, as a function of temperature, compared with lattice results, divided by $N_c^2=3^2$, for fixed values of the magnetic field. The red, blue and green lines correspond to ${\cal B}=0.01 \, {\rm GeV}^2$,  ${\cal B}=0.1 \, {\rm GeV}^2$ and , ${\cal B}=0.2 \, {\rm GeV}^2$ respectively. The lattice results (black dots with error bars) were obtained in the limit ${\cal B} \to 0$. } \label{Fig:MagnetisationLattice}
\end{figure}

Next, in figure \ref{Fig:MagnetisationLattice}, we compare our results for the magnetisation density $M_r$ and the magnetic susceptibility $\chi_r$ of a magnetised conformal plasma against those obtained in lattice QCD for a magnetised quark-gluon plasma \cite{Bali:2014kia}. The left panel of figure \ref{Fig:MagnetisationLattice} shows the magnetisation density, divided by $N_c^2$, as a function of the temperature for fixed values of the magnetic field. The red, blue, green and gray lines correspond to ${\cal B}=0 \, {\rm GeV}^2$, ${\cal B}=0.2 \, {\rm GeV}^2$, ${\cal B}=0.3 \, {\rm GeV}^2$ and ${\cal B}=0.4 \, {\rm GeV}^2$. The solid lines represent our results whilst the dots with error bars correspond to lattice QCD results (the latter divided by $N_c^2=3^2$). In the regime of high temperature the magnetisation density of the conformal plasma and the quark-gluon plasma show a similar increasing behaviour. However,  in the low temperature regime the results for the conformal plasma differ significantly from the lattice QCD results. The reason for this is that conformal symmetry breaking, chiral symmetry breaking and confinement play an important role in this regime and those effects are absent in the conformal plasma. We also note that the magnetisation density in the conformal plasma goes to zero in the limit of zero temperature. This is a consequence of our renormalisation procedure where we subtracted the full zero temperature result to eliminate the scheme dependence. This procedure is not exactly the same as that used in lattice QCD, where only the quadratic term of the zero temperature result is subtracted \cite{Bali:2014kia}. The right panel of figure \ref{Fig:MagnetisationLattice} shows our results for the magnetic susceptibility (solid lines), divided by $N_c^2$, as a function of the temperature for fixed values of the magnetic field. The red, blue and green lines correspond to ${\cal B}=0.01 \, {\rm GeV}^2$,  ${\cal B}=0.1 \, {\rm GeV}^2$ and , ${\cal B}=0.2 \, {\rm GeV}^2$ respectively.  This is compared with the lattice QCD results for the magnetic susceptibility (black dots with error bars), divided by $N_c^2=3^2$,  obtained in the limit of zero magnetic field \cite{Bali:2014kia}. In the high temperature regime the results for the conformal plasma are qualitatively similar to the lattice QCD results showing a monotonically increasing behaviour. At low temperatures the magnetic susceptibility goes to zero whilst the magnetic susceptibility found in lattice QCD becomes negative. Again, the differences between the results for the conformal plasma and the lattice QCD results are expected due to the fact that conformal symmetry breaking, chiral symmetry breaking and confinement are dominant effects in this regime. 

\begin{figure}[ht!]
\centering
\includegraphics[width=0.44\textwidth]{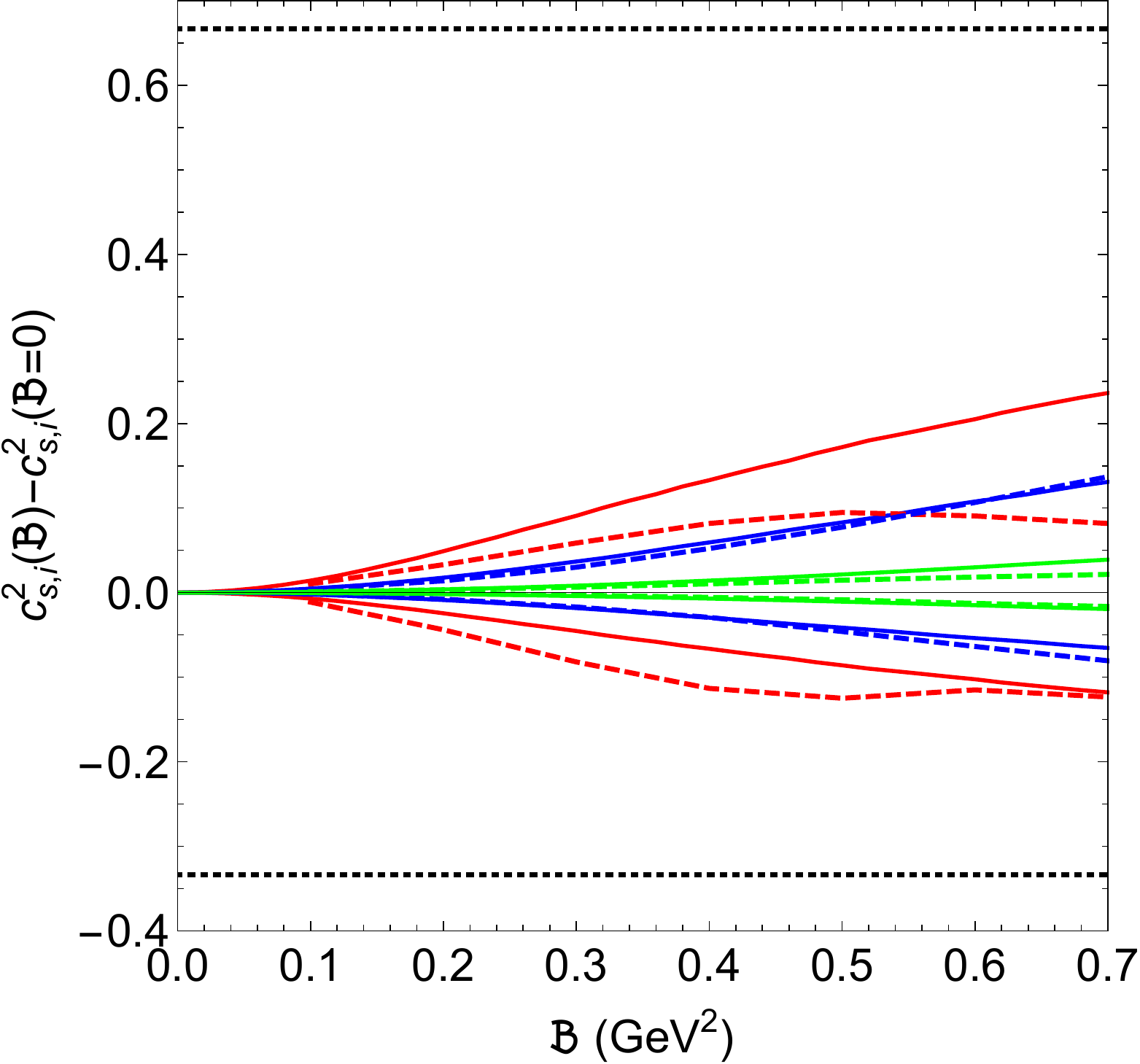}
\caption{Squared speeds of sound (solid lines) $c_{s,i}^2$ as functions of ${\cal B}$, with their values at ${\cal B}=0$ subtracted,  , compared to a numerical estimate obtained from lattice QCD results (dashed lines) for fixed values of the temperature. For simplicity, we omitted the error bars in the  numerical estimate. The lines above zero correspond to the longitudinal direction $i=z$ whereas the lines below zero correspond to the transverse direction $i=x$. The dotted upper and lower lines correspond to the values $2/3$ and $-1/3$ respectively, expected in the limit ${\cal B} \to \infty$. The red, blue and green colours correspond to $T = 0.15 \, \text{GeV}$, $T = 0.2 \, \text{GeV}$ and $T = 0.3 \, \text{GeV}$ respectively.}
\label{Fig:soundlattice}
\end{figure}

Lastly, in figure \ref{Fig:soundlattice} we present  our results for the variation of the squared speeds of sound in the longitudinal and transverse directions with the magnetic field (solid lines) compared with a numerical estimate that we made based on the lattice QCD results obtained in \cite{Bali:2014kia} (dashed lines), for fixed values of the temperature. The red, blue and green colours correspond to $T = 0.15 \, \text{GeV}$, $T = 0.2 \, \text{GeV}$ and $T = 0.3 \, \text{GeV}$ respectively.
 The variation of the squared speeds of sound is defined as  $c_{s,i}^2({\cal B}) - c_{s,i}^2({\cal B}=0)$. The lines above zero correspond to the longitudinal direction $i=z$ whereas the lines below zero correspond to the transverse direction $i=x$. Whilst the  variations of the squared speeds of sound in the conformal plasma always display a monotonic behaviour (increasing for $i=z$ and decreasing for $i=x$) the lattice QCD results display a similar behaviour only in the regime of high temperatures. As  explained above, we expect significant differences between the results for the conformal plasma and the lattice QCD results at low temperatures due to effects of conformal symmetry breaking, chiral symmetry breaking and confinement. 

\section{Conclusions}
\label{Sec:Conclusions}

In this paper we investigated, using the holographic framework, the thermodynamics of the strongly coupled ${\cal N}=4$ Super Yang-Mills plasma 
in the presence of a magnetic field. We introduced a holographic renormalisation procedure  that allowed us to obtain a Gibbs free energy and a stress-energy tensor consistent with the thermodynamics and hydrodynamics of a conformal fluid. Moreover,  this procedure allowed us to obtain scheme independent thermodynamic quantities that are consistent with the thermodynamics of the $BTZ\times \R^2$ solution at large values of the magnetic field. We also found a thermodynamic entropy consistent with the Bekenstein-Hawking entropy. 

We evaluated several thermodynamic quantities with an emphasis on  the anisotropic effects that the magnetic field induces. Our analysis is twofold: initially we performed a magnetohydrodynamic analysis to derive the equation of 
state and the stress-energy tensor of the conformal fluid and following this we introduced the magnetic brane solution, which interpolates between an  $AdS_5$ boundary 
and a $BTZ\times \R^2$ black hole in the deep IR.  
We implemented a holographic renormalisation procedure that  allowed us to find a Gibbs free energy and a stress-energy tensor consistent with the thermodynamics and hydrodynamics of a conformal fluid. The renormalisation procedure  included a subtraction of the action with the zero temperature result. This ensured that the end of the RG flow is given by the $BTZ\times \R^2$ solution and allowed us to obtain thermodynamic quantities that are scheme independent.

Regarding the conformal anomaly due to the presence of the magnetic field when considering a non-diffeomorphism invariant counterterm we conclude that, although the field theory deserves further investigation, our results suggest that it does not affect the thermodynamics and the, $0^{\rm th}$ order, hydrodynamics  of the fluid. Moreover, we have provided a prescription for removing the conformal anomaly from the very beginning by considering a diffemorphism invariant counterterm.

Having obtained the renormalised action, we performed both analytical (for small and large values of the dimensionless
parameter $b={\cal B}/ T^2$) and numerical computations of thermodynamic quantities. We note that, besides the dimensionless specific heat $C_{V,{\cal B}}/T^3$,   all of these thermodynamic quantities display a monotonic behaviour with $b$. The presence of the anisotropy due to the magnetic field leads to differing transverse and longitudinal pressures as well as the corresponding speeds of sound. This anisotropy in the hydrodynamic quantities shows up as a renormalisation flow from isotropy in the limit of the 4d conformal plasma to the anisotropy in the IR. This is most clearly seen in figure \ref{fig:RenPressures}. Close to the $AdS_5$ boundary (corresponding to small $b$) the conformal plasma is isotropic, but as we move towards the IR (corresponding to large $b$) the anisotropy becomes significant and the longitudinal pressure gets much higher than the transverse pressure. The plot for the two
speeds of sound in figure  \ref{fig:cs2} was even more illuminating, both because of the induced anisotropy and of the RG flow that the 
gravity solution depicts. As $b$ increases the squared longitudinal speed of sound grows from $1/3$ to $1$ whereas the squared transverse speed of sound decreases from $1/3$ to $0$. This behaviour is consistent with an RG flow between a 4d CFT at 
small $b$ to a 2d CFT at large $b$. With the above results being said, we note a caveat whereby a thermodynamic instability appears at around ${\cal B}= 12 T^2$.

In addition to our analytical and numerical analysis of magnetohydrodynamic quantities, we presented a phenomenological comparison between a number of these quantities for the 
magnetised conformal plasma and the lattice QCD results for a magnetised quark-gluon plasma.
As far as the hydrodynamic pressures are concerned the main conclusion is that the anisotropy between the two pressures in a magnetised conformal plasma and in the QCD quark-gluon plasma
increases with the magnetic field in a qualitatively similar way. For the speeds of sound whilst the variations of the squared speeds of sound in the conformal plasma always display a monotonic behaviour, the lattice QCD results display a similar behaviour only in the regime of high temperatures, as expected from asymptotic freedom.

Moving forward, an important set of phenomena of interest in QCD in the presence of a magnetic field
are those of magnetic catalysis and inverse magnetic catalysis. 
They are characterised by the strengthening and weakening of the chiral condensate, respectively in the presence of a magnetic field. 
The magnitudes of the magnetic field and the temperature, with respect to the deconfinement scale, determine which one of these will be realised. 
 
Whilst the physical mechanism behind MC is well understood, that of IMC remains an open problem. 
A promising explanation that has come from the lattice perspective \cite{Bruckmann:2013oba}, 
but it is also supported by arguments from holography \cite{Gursoy:2016ofp},  is the following:   
In the path integral, there is a competition between the trace of the inverse of the Dirac operator (the {\it valence} contribution) and the quark determinant (the {\it sea} contribution). Close to the deconfinement temperature, due the intricate dependence of the determinant on $B$ and $T$, the condensate is suppressed. 
In the holographic manifestation of those ideas, the {\it valence} contribution is from the tachyon field equation
and the  {\it sea} contribution is from the backreaction of the magnetic field on the gravitational  background. The
important ingredient for IMC is exactly this presence of the backreaction.

The bottom line is that whilst the realisation of MC is straightforward in a holographic 
framework, IMC is much harder to engineer.  The approach of \cite{Gursoy:2016ofp} is based on the sophisticated 
 Veneziano-QCD \cite{Jarvinen:2011qe} (V-QCD) holographic model. It is a combination of the  improved 
holographic QCD for the gluon sector \cite{Gursoy:2007er,Gursoy:2007cb} and the tachyon-DBI for the quark sector 
\cite{Bigazzi:2005md,Sen:2003tm}. It would be very interesting to investigate whether the emergence of IMC
is related to the complexity of V-QCD or whether a simpler gravity model that contains only the essential ingredients 
could also realise IMC.  In particular, it would be desirable to build a holographic QCD model that not only describes MC and IMC but also reduces to the anisotropic magnetic black brane solution of D'Hoker and Kraus in some limit. In fact, there have been some interesting developments for describing IMC starting with the D'Hoker and Kraus solution \cite{Mamo:2015dea,Dudal:2015wfn,Dudal:2016joz}
in the small $b$ regime where the solution is analytical. 

Hence, the natural extension of the current paper is the addition of flavor through a tachyon-DBI (with predetermined potentials) 
that will backreact on the colour dynamics. This may or may not be enough to realise IMC. 
The simplicity of the aforementioned construction will allow us to consider the effect of both 
massless and massive quarks on the potential appearance of IMC.    

A very interesting extension of these holographic ideas  
is claiming that the cause for IMC is not the charge dynamics that the magnetic field creates, but rather the anisotropy 
that it induces \cite{Giataganas:2017koz,Gursoy:2018ydr} . It would be very interesting to further test those ideas in the D'Hoker and Kraus framework (or in 4d using, as a starting point, the Hartnoll-Kovtun solution \cite{Hartnoll:2007ai}), either by replacing the 
magnetic field with an anisotropy or, in a more complex setup, by combining anisotropy and a magnetic field to further explore 
the competition and interplay between them (in this direction see also  \cite{Gursoy:2020kjd,Arefeva:2020vae}).

We would like to finish this paper stressing the importance of investigating QCD in the presence of magnetic field. This is relevant to the phenomenology of heavy ion collisions at non-zero impact parameter where the strong and time dependent magnetic field generated by the charged spectators induces electric currents in the quark-gluon plasma \cite{Gursoy:2014aka,Chatterjee:2018lsx}. These currents have a significant contribution to the so called ``directed flow'' coefficient $v_1$ in the hadronic spectrum which is being tested in Au$+$Au collisions at RHIC \cite{STAR:2019clv}. Moreover, recently a new window in the study of QCD at high density has opened, through advances coming from astrophysical observations
 of compact stars. Constraints on the masses and radii of these stars coming from LIGO, Virgo and the 
X-ray telescope NICER provide information on the equation of state of dense matter \cite{LIGOScientific:2018cki}. There is also an ongoing search for gravitational wave signals in the case of compact stars with strong magnetic fields, the so called magnetars \cite{LIGOScientific:2019ccu}.

\section*{Numerical code available}
We have made the code for the numerical calculations available within this work at\newline \href{https://www.wolframcloud.com/obj/jon.shock/Published/Numerics.nb}{https://www.wolframcloud.com/obj/jon.shock/Published/Numerics.nb}.


\section*{Acknowledgments}
The work of A.B-B is partially funded by Conselho Nacional de Desenvolvimento Cient\'\i fico e Tecnol\'ogico (CNPq, Brazil), grant No. 314000/2021-6, and Coordena\c{c}\~ao de Aperfei\c{c}oamento do Pessoal de N\'ivel Superior (CAPES, Brazil), Finance Code 001.
The work of D.Z has received funding from the Hellenic Foundation for Research and Innovation (HFRI) and the General Secretariat for Research and Technology (GSRT), under grant agreement No 15425.


\appendix 

\section{The perturbative solution at small ${\cal B}/T^2$}
\label{Appendix:smallb}

In this section, we will study the system perturbatively in powers of $B$ around the analytically known solution for $B$ = 0. 
This expansion is expected to be reliable for $B/T^2 \ll 1$.
To perform such an analysis we have to solve the usual set of equations of motion \eqref{fieldeqsv2} in this limit. In section 6 of \cite{DHoker:2009ixq} there are semi-analytical expressions for the different metric functions in 
the presence of charge density in that limit. However, if we restrict to the magnetic field the perturbative 
equations for the metric functions can be solved analytically. In order to be self contained we will repeat this analysis here and then we will use those results to study the thermodynamics. 

The functions $U$, $V$ \& $W$ are even in $B$ and we expand them up to order $B^2$
\begin{equation}
U(r) = r^2 -\frac{r_h^4}{r^2} + B^2\, U_2(r)\, , \quad V(r) = \log r + B^2\, V_2(r) \quad \& \quad 
W(r) = \log r + B^2\, W_2(r) \, . 
\end{equation}
As in \cite{DHoker:2009ixq} we introduce the following combinations
\begin{equation}
S_2 \, = \, 2\, V_2 + W_2 \quad \& \quad T_2 \, = \, V_2 -W_2
\end{equation}
to diagonalise the system of equations we have to solve. The equation for $T_2$ becomes
\begin{equation}
\partial_r \left[ r^3 \, \left(r^2 -\frac{r_h^4}{r^2} \right) T_2' \right] + \frac{2}{r}\, = \,0 
\end{equation}
and imposing smoothness for the first derivative of $T_2$ at the horizon 
and the vanishing of $T_2$ at infinity the solution is
\begin{equation}
T_2(r) \, = \, \frac{1}{r_h^4} \Bigg[\frac{\pi^2}{48} \, +\, \ln^2\left(\frac{r}{r_h}\right)\, + \, \frac{1}{8} \, \mathrm{Li}_2 
\left[1 -\frac{r^4}{r_h^4} \right]  \Bigg] \, .
\end{equation}
The equation for $S_2$ and the corresponding solution is  
\begin{equation}
\partial_r \left[ r^3 \, S_2' \right] \, = \,0 \quad \Rightarrow \quad S_2(r) = 0 \, ,
\end{equation}
since we need the expression for $S_2$ to fall off faster than $1/r^2$ at infinity. Finally, the equation for $U_2$ becomes
\begin{equation}
\partial_r \left[ r^3 \, U_2' \right] - \frac{4}{3\, r}\, = \,0 \, . 
\end{equation}
The solution of the equation above is the following
\begin{equation}
U_2 (r) \, = \, - \, \frac{2 \ln r +1 + \gamma_3}{3\, r^2}\, , 
\end{equation}
where the first integration constant was fixed by the requirement that $U_2$ vanishes at infinity and  
$\gamma_3$ is the second (undetermined) integration constant. To summarise the analysis so far, the perturbative solution 
of the equations of motion in \eqref{fieldeqsv2}  up to order $B^2$ is the following
\begin{eqnarray} \label{small_B_solution}
U(r) & = &  r^2 -\frac{r_h^4}{r^2} - B^2\, \frac{2 \ln r +1 + \gamma_3}{3\, r^2}
\nonumber\\
V(r) &= & \log r + \frac{1}{3} \, \frac{B^2}{r_h^4} \Bigg[\frac{\pi^2}{48} \, +\, \ln^2\left(\frac{r}{r_h}\right)\, + \, \frac{1}{8} \, \mathrm{Li}_2 
\left[1 -\frac{r^4}{r_h^4} \right]  \Bigg] 
\\
W(r) &= & \log r - \frac{2}{3}\, \frac{B^2}{r_h^4} \Bigg[\frac{\pi^2}{48} \, +\, \ln^2\left(\frac{r}{r_h}\right)\, + \, \frac{1}{8} \, \mathrm{Li}_2 
\left[1 -\frac{r^4}{r_h^4} \right]  \Bigg] \, .  
\nonumber
\end{eqnarray}
The constant $\gamma_3$ is fixed to 
\begin{equation}  \label{small_B_gamma3}
\gamma_3 \, = \, - \, 1 - 2 \, \ln r_h \, , 
\end{equation}
by the requirement $U(r_h)=0$ for the horizon function.
Using the above expressions we can calculate the temperature and the entropy density in the usual fashion 
\begin{equation} \label{small_B_temperature}
T \, = \, \frac{U'(r_h)}{4 \pi} \quad \Rightarrow \quad 
T \, =\, \frac{r_h}{\pi} \left[1 - \frac{1}{6} \, \frac{B^2}{r_h^4} \right]
\end{equation}
and
\begin{equation} \label{small_B_entropy_v1}
{\cal S} \equiv \, \frac{A_h}{4\, G_5 \, V_3} \, = \, \frac{N_c^2}{2\, \pi}\, e^{W(r_h)+2\,V(r_h)} \quad \Rightarrow \quad 
{\cal S} \, = \,N_c^2 \,  \frac{r_h^3}{2\, \pi} \, . 
\end{equation}
Notice that whilst there is a correction for the temperature at the order $B^2$, there is no correction for the entropy. 


\subsection{Free energy and Magnetisation}

Next step in the analysis will be the calculation of the on-shell action (and correspondingly of the free energy). 
The Einstein-Maxwell action is given by the following expression 
\begin{equation}
S \, = \, \frac{1}{16 \, \pi \, G_5} \int d^5 x \, \sqrt{-g} \, \Big[R +12 - F^2 \Big]
\quad {\rm with} \quad F^2 \equiv F_{\mu \nu} \, F^{\mu \nu} 
\end{equation}
and substituting the perturbative analytical solution of \eqref{small_B_solution} the action integral becomes
\begin{equation} \label{small_B_action}
\frac{8 \, \pi \, G_5}{V_3\, \beta } \, S\, = \, -\, \int_{r_h}^{\Lambda} \, dr  \left[4 \, r^3 + \frac{2}{3} \, \frac{B^2}{r}\right] \, 
= \, r_h^4 - \Lambda^4 -\frac{2}{3}\, B^2 \, \ln \left(\frac{\Lambda}{r_h}\right)
\end{equation}
where $\beta$ is the inverse temperature and $\Lambda$ is the UV cut-off. 
The sum of the Gibbons-Hawking boundary term $S_{\partial {\cal M}}$ and the counterterm $S_{ct}$, needed  to cancel the UV divergences, is given by the following expression (see also \eqref{GK5d} and \eqref{counterterm-action})
\begin{equation}
S_{\partial {\cal M}} + S_{ct} \, = \, \frac{1}{8 \, \pi \, G_5} \int d^4 x \, \sqrt{-\gamma} \, \Big[K -3 - 
\frac{1}{8} \, F^2 \, \ln F^2 - \frac{a_3}{2} \, F^2 \Big]
\end{equation}
and after substituting the perturbative solution \eqref{small_B_solution} it becomes
\begin{equation} \label{small_B_ct}
\frac{8 \, \pi \, G_5}{V_3\, \beta } \, \Big ( S_{\partial {\cal M}} + S_{ct} \Big )\, = \, \Lambda^4 - \frac{r_h^4}{2} + B^2 \left[\frac{1}{12}\, 
\ln \left(\frac{ \Lambda^8}{ 8\,B^6}\right) - \frac{1}{2} - \frac{\gamma_3}{6}-a_3 \right] \, . 
\end{equation}
Summing equations \eqref{small_B_action} and \eqref{small_B_ct} we calculate the renormalised action
\begin{equation} \label{small_B_ren}
\frac{8 \, \pi \, G_5}{V_3\, \beta } \, S_{ren}\, = \, \frac{r_h^4}{2} + B^2 \left[\frac{1}{12}\, 
\ln \left(\frac{ r_h^8}{8\, B^6}\right) - \frac{1}{2} - \frac{\gamma_3}{6} - a_3\right] \, . 
\end{equation}
The renormalised (Gibbs) free energy density takes the following form
\begin{equation} \label{small_B_free_energy_v1}
{\cal G} \, = \, - \frac{S_{ren}}{V_3 \, \beta} \quad \Rightarrow \quad {\cal G} \, =\, - \frac{N_c^2}{4 \, \pi^2} \, \Bigg[\frac{r_h^4}{2} + B^2 
\left[\frac{1}{12}\, \ln \left(\frac{ r_h^8}{8\, B^6}\right) - \frac{1}{2} - \frac{\gamma_3}{6} - a_3\right] \Bigg] \, . 
\end{equation}
Using \eqref{small_B_temperature} and  \eqref{small_B_free_energy_v1} we can differentiate the free energy with respect to the temperature and calculate the entropy. In this way we can identify the two different (but complementary) computations for the entropy: the first one coming from the first law of thermodynamics and 
the second one from the Bekenstein-Hawking formula \eqref{small_B_entropy_v1}. 

Using the value for $\gamma_3$ from \eqref{small_B_gamma3} the expression for the free energy density becomes
\begin{equation} \label{small_B_free_energy_v2}
{\cal G} \, =\, - \frac{N_c^2}{4 \, \pi^2} \, \frac{r_h^4}{2} \, \Bigg[1- \frac{B^2}{r_h^4} \left[\frac{1}{2}\, 
\ln \left(\frac{2\, B^2}{r_h^4} \right) + \frac{2}{3}+ 2\, a_3 \right] \Bigg] \, . 
\end{equation}
Following the reasoning we fully described in the main body of the text, we will subtract the free energy density at zero
temperature. From \eqref{free-energy-norm-zeroT} we have 
\begin{equation}  \label{free-energy-norm-zeroT_v2}
{\cal G}_{T=0} \,= \, - \,\frac{N_c^2 \, B^2}{8 \, \pi^2} \left [\tilde U_{\infty,4}(\sqrt{3}) -  \ln (\sqrt{3}) \right] + 
\frac{N_c^2 \,B^2}{4\, \pi^2} \left(a_3 + \frac{1}{4} \ln 2\right) \, .
\end{equation}
Subtracting \eqref{free-energy-norm-zeroT_v2} from \eqref{small_B_free_energy_v2} we obtain the following expression 
for the renormalised free energy
\begin{equation} \label{small_B_free_energy_v3}
{\cal G}_r \, =\, - \frac{N_c^2}{4 \, \pi^2} \, \frac{r_h^4}{2} \, \Bigg[1- \frac{B^2}{r_h^4} \left[\frac{1}{2}\, 
\ln \left(\frac{B^2}{3\, r_h^4} \right) + \frac{2}{3}+ \tilde U_{\infty,4}(\sqrt{3})  \right] \Bigg] \, . 
\end{equation}
This is the free energy we will use in the rest of the analysis to calculate all the thermodynamic quantities.
Using \eqref{small_B_temperature}, \eqref{small_B_free_energy_v3}
and the well known relation between the physical magnetic field $\cal B$ and the parameter $B$ 
(namely ${\cal B} = \sqrt{3} \,B$) we can express the free energy as a function of the temperature and 
the dimensionless ratio ${\cal B}/T^2$. It reads
\begin{equation} \label{small_B_free_energy_over-T^4}
\frac{{\cal G}_r}{N_c^2 \, T^4} \, =\, - \frac{\pi^2}{8} + \frac{1}{24 \, \pi^2} \, \frac{{\cal B}^2}{T^4} 
\Bigg[\tilde U_{\infty,4}(\sqrt{3})  + \ln \left( \frac{1}{3\, \pi^2}\, \frac{{\cal B}}{T^2} \right) \Bigg] \, . 
\end{equation}
The thermodynamic entropy density is obtained from the first law of thermodynamics. Namely, we use the relation ${\cal S} = - \partial {\cal G}_r/ \partial T$ and find 
\begin{equation} \label{small_B_entropy_over-T^3}
\frac{{\cal S} }{N_c^2 \, T^3} \, = \, \frac{\pi^2}{2} \left(1+ \frac{1}{6 \, \pi^4} \, \frac{{\cal B}^2}{T^4} \right) \, . 
\end{equation}
This agrees with the Bekenstein-Hawking entropy density \eqref{small_B_entropy_v1}, which is a non-trivial check of the renormalisation procedure. Now that we have fully specified the expression for the free energy density we can calculate the magnetisation density
using the following formula
\begin{equation} \label{small_B_magnetisation}
M_r \, = \, - \, \partial_{\cal B} {\cal G}_r  \quad \Rightarrow \quad
\frac{M_r}{N_c^2 \, T^2} \, = \, - \frac{1}{12 \, \pi^2} \, \frac{{\cal B}}{T^2} \, \left[\tilde U_{\infty,4}(\sqrt{3})  + \frac{1}{2} 
+ \ln \left(\frac{1}{3 \, \pi^2}  \frac{{\cal B}}{T^2} \right)\right] \, . 
\end{equation}
Combining the expressions for the entropy, temperature and magnetisation we can write the expression of the free energy density solely in terms of field theory quantities
\begin{equation}  \label{small_B_free_energy_v4}
{\cal G}_r \, = \, - \, \frac{1}{4} \, {\cal S} \, T \, - \, \frac{1}{2}\, {\cal B} \, M_r \, . 
\end{equation}
This is exactly the expression for the Gibbs free energy that was computed in the main text using the different 
approach of conformal magnetohydrodynamics. The same result was also obtained is a different holographic model 
in \cite{Ammon:2012qs}. That was a top-down construction with D3 branes, a large number of backreacted smeared 
D7 branes, temperature and a magnetic field. The solution was perturbative in the backreaction parameter and the 
magnetic field was sourcing an anisotropy. All the thermodynamic quantities were computed at first order in the 
backreaction parameter and relation \eqref{small_B_free_energy_v4} was verified. Notice that it makes 
complete sense that the on-shell action we calculate from holography is related to the Gibbs free energy, since the 
computation is performed in an ensemble with fixed magnetic field. 


\subsection{Stress Energy Tensor}

The calculation of the different components of the energy momentum tensor is presented in full detail 
in the main text. Here we will substitute on those expressions the perturbative analytical solution of \eqref{small_B_solution} 
and focus the attention on the vev of the SE tensor
\begin{equation}
\langle T^i_{\,\,j} \rangle \, = \, \sqrt{- \gamma} \, T^i_{\,\,j} \, ,
\end{equation}
since the components of the vev are related to the transverse and the longitudinal pressures. 
An important comment is in order: In order to construct a solution that flows from $AdS_5$ on the boundary to 
$\rm \bf BTZ \times \R^2$ in the IR, we performed two consecutive holographic renormalisations of the initial action. In the first renormalisation we added the covariant counterterms and in the second renormalisation we subtracted the zero temperature result.  The second step is equivalent to fixing the $a_3$ coefficient of the counterterm action \eqref{counterterm-action} to the following  value
\begin{equation}  \label{fixing_a3}
a_3 \, = \, \frac{1}{2} \,  \tilde U_{\infty,4}(\sqrt{3})  - \frac{1}{4} \, \ln 6 \, . 
\end{equation}
After that procedure, we can substitute the perturbative analytical solution of \eqref{small_B_solution} 
on the general expressions for the components of the energy momentum tensor. For the $tt$ component we have
\begin{equation} \label{small_B_SE_tt}
\langle T^t_{\,\,t} \rangle = - \, \frac{3 \, r_h^4}{8\, \pi^2} 
\, N_c^2 \,
\Bigg[1 + \frac{B^2}{6\, r_h^4}\left[2 \, \tilde U_{\infty,4}(\sqrt{3})  + 
\ln \left(\frac{B^2}{3\, r_h^4}\right)\right] \Bigg] 
\,\, \Rightarrow \,\,
\langle T^t_{\,\,t} \rangle = - \frac{3}{4} \, {\cal S} \, T + \frac{1}{2} \, M \,  {\cal B}
\end{equation}
where in the last step we have expressed everything in terms of field theory quantities. 
The result we obtain is minus the energy density and agrees with the computation that is detailed in the main text 
from the conformal magnetohydrodynamics approach. For the rest of the components of the SE tensor 
we obtain the following results
\begin{eqnarray} \label{small_B_SE_11_22_v1}
&& \langle T^1_{\,\,1} \rangle \, = \, \langle T^2_{\,\,2} \rangle\,  \, = \,\frac{1}{4} \, {\cal S} \, T \, - \frac{1}{2} \, M \,  {\cal B} \, 
\equiv \, P_{x} \\ \label{small_B_SE_33_v1}
&& \langle T^3_{\,\,3} \rangle  \, = \,\frac{1}{4} \, {\cal S} \, T \, + \frac{1}{2} \, M \,  {\cal B} \, = \,  \langle T^1_{\,\,1} \rangle \, 
+ \,  M \,  {\cal B}  
\equiv \, P_{z} \, . 
\end{eqnarray}
The transverse and longitudinal pressures as functions of the temperature and ${\cal B}/T^2$ read
\begin{eqnarray} \label{small_B_SE_11_22_v2}
&&  \frac{P_{x}}{N_c^2 \, T^4} \, = \, \frac{\pi^2}{8} \Bigg[1+ \frac{1}{3\, \pi^4}\,\frac{{\cal B}^2}{T^4} \left[1+  
\tilde U_{\infty,4}(\sqrt{3})  
+ \ln \left(\frac{1}{3 \, \pi^2}  \frac{{\cal B}}{T^2} \right)\right]\Bigg]
\\  \label{small_B_SE_33_v2}
&& \frac{P_{z}}{N_c^2 \, T^4}  \, = \, \frac{\pi^2}{8} \Bigg[1- \frac{1}{3\, \pi^4}\,\frac{{\cal B}^2}{T^4} \left[ 
\tilde U_{\infty,4}(\sqrt{3})  + 
\ln \left(\frac{1}{3 \, \pi^2}  \frac{{\cal B}}{T^2} \right)\right]\Bigg] \, . 
\end{eqnarray}
The results for the hydrodynamic pressures obtained above were obtained using the holographic dictionary for the stress-energy tensor, described in \ref{subsec:Hologstress}, and is consistent with the identification of the longitudinal pressure $P_z$  with minus the Gibbs free energy density, i.e. $P_z = - {\cal G}$, and the identification of the transverse pressure $P_x$ with minus the Helmoltz free energy density, i.e. $P_x = - {\cal F} = -{\cal G} - M {\cal B}$. These relations were expected for a magnetised conformal plasma, as described in section \ref{Sec:CFT}. This is 
a non-trivial consistency check of the holographic renormalisation procedure and it is also consistent with \cite{Mateos:2011tv,Ammon:2012qs}.

From \eqref{small_B_SE_11_22_v2}, which according to the identification we already discussed corresponds to minus the Helmholtz free energy, and \eqref{small_B_magnetisation} we can verify the following thermodynamic relations 
\begin{equation}
{\cal S} = -  \frac{\partial F_H}{\partial T}\Bigg \vert_M \quad \& \quad 
{\cal B} \, = \, \frac{\partial F_H}{\partial M}\Bigg \vert_T 
\quad \text{with} \quad F_H = - P_{x} \, .
\end{equation}
More specifically, to work at fixed magnetisation we have to specify the way that $B$ changes with the temperature. 
For that we have to consider that $B$ in \eqref{small_B_magnetisation} depends on temperature and 
set the derivative with respect to $T$ to zero. Solving this equation we obtain the following relation 
\begin{equation}
\partial_T {\cal B} \, = \, \frac{4 \, {\cal B}}{T} \, \Bigg[3+ 2 \, \tilde U_{\infty,4}(\sqrt{3})  + 2 \, 
 \ln \left(\frac{1}{3 \, \pi^2}  \frac{{\cal B}}{T^2} \right) \Bigg]^{-1} \, . 
\end{equation}


\subsection{Susceptibilities and the Speed of Sound}

Taking the derivative of the magnetisation it is possible to calculate the two susceptibilities, pyro-magnetic 
($\xi$) and magnetic ($\chi$). 
In the majority of the holographic constructions the pyro-magnetic susceptibility is vanishing, but in this model due to the 
backreaction it is not. We obtain the following results 
\begin{equation}  \label{small_B_xi_chi}
\frac{\xi}{N_c^2 \, T} \, = \, \frac{1}{6 \, \pi^2} \,  \frac{{\cal B}}{T^2}
\quad \& \quad
 \frac{\chi}{N_c^2}  =  - \frac{1}{24 \, \pi^2} \Bigg[3 +2 \, \tilde U_{\infty,4}(\sqrt{3}) + 
2\, \ln \left(\frac{1}{3 \, \pi^2}  \frac{{\cal B}}{T^2} \right) \Bigg] \, . 
\end{equation}
Using \eqref{small_B_entropy_over-T^3}, \eqref{small_B_magnetisation} and \eqref{small_B_xi_chi}  
it is easy to confirm that the relation between magnetisation, temperature and the  two susceptibilities that is dictated by conformal magnetohydrodynamics in \eqref{conformal_identity}, is satisfied up to second order in $\cal B$. 

Finally we will analyse the speed of sound in the magnetic plasma. Since there is an
anisotropy in the gravity solution, there two  directions for the pressure waves to propagate, each one with a
different speed. For a perturbation along the direction of the magnetic field we have
\begin{equation} \label{small_B_speed_parallel_v1}
c^2_{s, z} \, = \, \frac{\partial P_{z}}{\partial \rho} \, = \, \frac{- \frac{\partial {\cal G}}{\partial T}\big \vert_B}{\frac{\partial {\rho}}{\partial T}\big \vert_B} \, = \, \frac{\cal S}{C_{V,{\cal B}}}
\end{equation}
where $\rho$ is the magnetic enthalpy density and $C_{V,{\cal B}}$ is the specific heat at fixed magnetic field.
At second order in $\cal B$ the specific heat is given by the following expression
\begin{equation} \label{small_B_heat_capacity}
\frac{C_{V,{\cal B}}}{N_c^2 \, T^3} \, = \, \frac{3 \, \pi^2}{2} \left[1 - \frac{1}{18 \, \pi^4} \,\frac{{\cal B}^2}{T^4} \right] \, . 
\end{equation}
Plugging \eqref{small_B_heat_capacity} into \eqref{small_B_speed_parallel_v1} we can calculate the speed of sound in the 
direction of the magnetic field
\begin{equation} 
c^2_{s, z} \, = \,  \frac{1}{3} \, \left[ 1 + \frac{2}{9\, \pi^4} \, \frac{{\cal B}^2}{T^4}\right]
\label{cs2zsmallb}
\end{equation}
which is above the conformal result at zero magnetic field. 

Moving to the direction orthogonal to the magnetic field, the speed of sound is given by the following expression 
\begin{equation} 
c^2_{s, x} \, = \, \frac{\partial P_{x}}{\partial \rho} \, = \, \frac{\cal S}{C_{V,{\cal B}}} -  \, \frac{B}{C_{V,{\cal B}}} \, 
\frac{\partial M}{\partial T}\Bigg \vert_B \, . 
\end{equation}
Plugging the expressions for the different quantities up to second order in $\cal B$ we obtain the following result
\begin{equation} 
c^2_{s, x} \, = \,  \frac{1}{3} \, \left[ 1 - \frac{1}{9\, \pi^4} \, \frac{{\cal B}^2}{T^4}\right]
\label{cs2xsmallb}
\end{equation}
which is below the conformal result at zero magnetic field.


\section{The $\rm \bf BTZ \times \R^2$ solution at large ${\cal B}/T^2$}
\label{Appendix:BTZ}

In this section we will present the solution of the equations of motion and elaborate on the thermodynamics 
at large values of ${\cal B}/T^2$. 
Contrary to the case of small ${\cal B}/T^2$ that we analysed in the previous section of the appendix, here we have 
not succeeded in expanding perturbatively around the BTZ solution. However, even without subleading terms we will 
compute the leading behavior for all of thermodynamic quantities at the end of the RG flow, 
that connects a 3+1 CFT at short distance with an 1+1 CFT at long distance. 

Using the ansatz for the background metric and the magnetic field given in \eqref{ansatz}, we look for a solution that is the product of a BTZ black hole \cite{Banados:1992wn} (in $t$, $z$ and $r$)  and $\R^2$. Hence, we consider the ansatz
\begin{equation}
V(r) = \text{constant} = V_0 \, . 
\end{equation}
Using this ansatz, the Einstein-Maxwell equations reduce to 
\begin{align}
& 2 \, B^2 \, e^{-4 \, V_0} - U' \, W' = 0  \label{EMBTZ1}  \\
& \left(W'\right)^2 + W'' = 0 \label{EMBTZ2} \\[4pt]
& - 12 + 2 \, B^2 e^{-4 \, V_0} + U' \, W' = 0 \label{EMBTZ3} \,. 
\end{align}
We remind that reader that the magnetic field ${\cal B}$ of the dual field theory is related to $B$ by  $B = {\cal B}/\sqrt{3}$, as described in \cite{DHoker:2009mmn}. 
Summing \eqref{EMBTZ1} and \eqref{EMBTZ3} we determine the constant $V_0$
\begin{equation}  \label{BTZconstr}
B^2 \, e^{-4 \, V_0} = 3 \quad \Rightarrow \quad V_0 \, = \, - \, \frac{1}{2} \, \ln \left(\frac{B}{\sqrt{3}}\right)\,.
\end{equation}
The most general solution to the differential equation \eqref{EMBTZ2} can be written as
\begin{equation}
W(r) = \ln a + \ln \left(r+\delta \right) \, .  \label{BTZWsol} 
\end{equation}
We fix $\delta=0$ because this integration constant simply represents a shift in the radial coordinate. 
In order to reproduce the $AdS_3$ asymptotics, with radius $\ell=1/\sqrt{3}$, of the BTZ solution we 
impose that $a= \sqrt{3}$.
Plugging \eqref{BTZconstr} into \eqref{EMBTZ1} or \eqref{EMBTZ3} we find the following differential equation for $U$
\begin{equation}
U' \, = \, 6\,  r  \quad \Rightarrow \quad U(r) \, = \, 3 \, r^2 + c \,. 
\end{equation}
The condition $U(r_h)=0$, with $r_h$ the horizon radius, allows us to fix $c= - 3 \, r_h^2$. 
The temperature is obtained from the usual formula
\begin{equation}
T \, = \, \frac{1}{4\pi} \, U'(r_h) = \frac{3\, r_h }{2\, \pi} \,. 
\end{equation}
The entropy density is given by the Bekenstein-Hawking area formula
\begin{equation}
{\cal S}= \frac{A_h}{4 \, G_5 \, V_3} = \frac{N_c^2}{2\, \pi}\, e^{2 \, V(r_h)+ W(r_h)} = 
\frac{N_c^2}{3 \, \sqrt{3}} \, {\cal B} \, T \, ,
\end{equation}
where we used the relation ${\cal B} = \sqrt{3} \, B$. The on-shell action can be written as
\begin{equation}
\frac{8 \, \pi \, G_5}{V_3\, \beta } \, S\, = \, 3 \, B \, \left( r_h^2 - \Lambda^2 \right) \, .
\end{equation}
The sum of the Gibbons-Hawking boundary term $S_{\partial {\cal M}}$ and the counterterm $S_{ct}$, needed  to cancel the UV divergence, is given by
\begin{align}
 &S_{\partial {\cal M}} +S_{ct} = \frac{1}{8 \, \pi \, G_5} \int d^4 x \, \sqrt{-\gamma} \, \left(K -a_1\right) \nonumber \\
\Rightarrow \,
& \frac{8 \, \pi \, G_5}{V_3\, \beta } \, \Big ( S_{\partial {\cal M}} +S_{ct} \Big )= B\, \left[\Lambda^2 - \frac{r_h^2}{2}\right] \left[6 - a_1 \, \sqrt{3}\right] \, . 
\end{align}
Note that there are no other possible counterterms in this case and in particular, there is no ambiguity due to the presence of finite counterterms. 
In order to cancel the UV divergences we choose $a_1 = \sqrt{3}$.  Then the renormalised action becomes
\begin{equation}
\frac{8 \, \pi \, G_5}{V_3\, \beta } \, S_{ren}\, = \, \frac{3}{2}\, B\, r_h^2 \, . 
\end{equation}
The renormalised free energy density becomes
\begin{equation}
{\cal G} =  \frac{S_{ren}}{V_3\, \beta}  \, = \, - \, \frac{3 N_c^2}{8\, \pi^2} \, B\, r_h^2 \, = \, - \, \frac{N_c^2}{6 \sqrt{3}} \, {\cal B} \, T^2 \, = \, - \frac{1}{2}\, T\, {\cal S}  \,. 
\label{FreeEnBTZ}
\end{equation}
The magnetisation density takes the form
\begin{equation}
M = -  \frac{\partial {\cal G}}{\partial {\cal B}} =  \frac{N_c^2}{6 \sqrt{3}} \, T^2 \, .
\end{equation}
The magnetic susceptibility $\chi$ vanishes and the pyro-magnetic coefficient becomes
\begin{equation}
\xi = \frac{ \partial M} { \partial T} = \frac{N_c^2}{3 \, \sqrt{3}} T \,.  
\end{equation}
Note that the Helmholtz free energy density vanishes and the internal energy density reduces to 
\begin{equation}
{\cal U}= T {\cal S} =   \frac{ N_c^2}{3 \, \sqrt{3}} {\cal B} \, T^2  \, .
\end{equation}
The enthalpy density takes the form
\begin{equation}
\rho = {\cal G} + T \, {\cal S} = \frac{1}{2} \, T \, {\cal S} \,.    
\end{equation}
The components of the stress-energy tensor become
\begin{equation}
\langle T^t_{\,\,t} \rangle  =  - \, \rho  \, , \quad
\langle T^1_{\,\,1} \rangle = \langle T^2_{\,\,2} \rangle = P_{x} = 0
\quad \& \quad 
\langle T^3_{\,\,3} \rangle = P_{z} = - \,  {\cal G} = \frac{1}{2} \, T \, {\cal S} \,. 
\end{equation}
Note that $ \rho - P_{z} = 0$, which is expected for a 2d CFT. The specific heat takes the form
\begin{equation}
 C_{V,{\cal B}} = \frac{ \partial {\cal \rho}}{\partial T}  =  {\cal S} \, .
\end{equation}
Finally, the squared speed of sound in the $x$ and $z$ direction reduce to
\begin{equation}
c_{s,x}^2 \, = \, \frac{ {\cal S} - \xi \, {\cal B}}{C_{V,{\cal B}}} \, = \,  0 \quad \& \quad 
c_{s,z}^2 \, = \, \frac{ {\cal S}}{C_{V,{\cal B}}} \, = \,  1 \, . \label{cs2xzlargeb}
\end{equation}


\section{Details on the calculation of the holographic stress tensor}
\label{Appendix:stress-tensor}

In this appendix we present details of the calculation of the holographic stress tensor that appears in subsection \ref{subsec:Hologstress}.
Substituting the ansatz \eqref{ansatz} in \eqref{Tmunu_Reg+ct} we obtain  
\begin{eqnarray}
&& \frac{T^{tt}_{reg}}{2 \, \sigma \, r_0^{6}} = \Bigg [ - \frac{2 \, V' + W'}{\sqrt{U}} \Bigg ]_{r_0}\, , \quad 
\frac{T^{11}_{reg}}{2 \, \sigma \, r_0^{6}} = \frac{T^{22}_{reg}}{2 \, \sigma \, r_0^{6}} = 
\Bigg [ \frac{U'+2\, U \left(V'+W'\right)}{2 \sqrt{U} \, e^{2V}} \Bigg ]_{r_0} 
\nonumber \\
&&
\quad \quad \& \quad \frac{T^{33}_{reg}}{2 \, \sigma \, r_0^{6}} = \Bigg [\frac{U'+4U V'}{2 \sqrt{U} \, e^{2W} } \Bigg ]_{r_0}
\end{eqnarray}
whilst \eqref{Tmunu_ct_v2} becomes
\begin{eqnarray}
&& \frac{T^{tt}_{ct}}{2 \, \sigma \, r_0^{6}} = 
\Bigg [ \frac{a_1 e^{4V} + 2 a_2 B^2 \ln (2 B^2 e^{-4V}) + 2 a_3 B^2  }{2\, U\, e^{4V}} \Bigg ]_{r_0}
\nonumber \\
&&
\frac{T^{11}_{reg}}{2 \, \sigma \, r_0^{6}} = \frac{T^{22}_{reg}}{2 \, \sigma \, r_0^{6}} = -
\Bigg [ \frac{e^{-6V}}{2} \Big [ a_1 e^{4V} - 2 a_2 B^2 \ln (2 B^2 e^{-4V}) - 2 (2 a_2 + a_3 ) B^2  \Big ] \Bigg ]_{r_0} 
\nonumber \\
&&
\quad \quad \& \quad \frac{T^{33}_{reg}}{2 \, \sigma \, r_0^{6}} = -
\Bigg [\frac{e^{-4V - 2W}}{2} \Big [ a_1 e^{4V} + 2 a_2 B^2 \ln (2 B^2 e^{-4V}) + 2 a_3 B^2 \Big ] \Bigg ]_{r_0} \,.
\end{eqnarray}
Plugging the UV asymptotic behaviour \eqref{UVNLO} and choosing $a_1=6$, $a_2=1/4$ we obtain
\begin{align}  \label{EMT_rho_v1}
\rho &= - \langle T^t_{\,\, t} \rangle = -\frac{r_h^4}{8\, \pi^2} \, N_c^2 \,\Bigg [ 3 \, \tilde U_{\infty,4} -  \tilde B^2 \, \ln \tilde B
 - \left(2 a_3 + \frac{1}{2} \, \ln 2 \right) \tilde B^2 \Bigg ] 
 \\  \label{EMT_P_x_v1}
P_x &= \langle T^1_{\,\, 1} \rangle= \langle T^2_{\,\, 2} \rangle= -
\frac{r_h^4}{8\, \pi^2} \, N_c^2 \, \Bigg [ \tilde U_{\infty,4} - 8 \, \tilde v_{\infty,4} - \tilde B^2 \ln \tilde B - 
\left(2 a_3 + \frac{1}{2} \, \ln 2 \right) \tilde B^2 \Bigg ] 
\\  \label{EMT_P_z_v1}
P_z &= \langle T^3_{\,\, 3} \rangle= -\frac{r_h^4}{8\, \pi^2} \, N_c^2 \,\Bigg [ \tilde U_{\infty,4} + 16 \, \tilde v_{\infty,4} 
+ \tilde B^2 \ln \tilde B + \left(2 a_3 + \frac{1}{2}\,  \ln 2\right) \tilde B^2 \Bigg ] \, . 
\end{align}
In the subsection \ref{subsubsec:Gibbs} we introduced a zero temperature subtraction in order to get rid of the scheme dependent 
parameter $a_3$. This subtraction is equivalent to fixing the $a_3$ coefficient by the expression found in \eqref{fixing_a3}.
In this case  \eqref{EMT_rho_v1}, \eqref{EMT_P_x_v1} \& \eqref{EMT_P_z_v1} are modified to obtain \eqref{EMT_rho_v2}, \eqref{EMT_P_x_v2} \&  \eqref{EMT_P_z_v2}.



\bibliographystyle{utphys}

\bibliography{MagPlasmav2}

\end{document}